\documentclass[10pt,a4paper,english,prd,nofootinbib,notitlepage,superscriptaddress,showkeys,preprintnumbers,longbibliography]{revtex4-1}
\usepackage[T1]{fontenc}
\usepackage[latin9]{inputenc}
\usepackage{color}
\usepackage{babel}
\usepackage{amsmath}
\usepackage{amssymb}
\usepackage{graphicx}
\usepackage{esint}
\usepackage[unicode=true,
 bookmarks=false,
 breaklinks=false,pdfborder={0 0 1},backref=false,colorlinks=true]
 {hyperref}
\hypersetup{
 citecolor=blue,urlcolor=blue,linkcolor=blue}
\usepackage{breakurl}

\makeatletter

 \@ifundefined{textcolor}{}
 {%
   \definecolor{BLACK}{gray}{0}
   \definecolor{WHITE}{gray}{1}
   \definecolor{RED}{rgb}{1,0,0}
   \definecolor{GREEN}{rgb}{0,1,0}
   \definecolor{BLUE}{rgb}{0,0,1}
   \definecolor{CYAN}{cmyk}{1,0,0,0}
   \definecolor{MAGENTA}{cmyk}{0,1,0,0}
   \definecolor{YELLOW}{cmyk}{0,0,1,0}
 }

\usepackage{babel}
\usepackage{babel}
\usepackage{babel}
\usepackage{babel}

\usepackage{babel}
\usepackage{breakurl}\usepackage{cleveref}\usepackage{ulem}\usepackage{bbm}

\usepackage{todonotes}
\definecolor{lightblue}{HTML}{A9D0F5}
\definecolor{lightgreen}{HTML}{BCF5A9}
\definecolor{lightred}{HTML}{F6CECE}
\definecolor{lightorange}{HTML}{FFA800}
\definecolor{greengray}{HTML}{5C9393}
\definecolor{lightgreengray}{HTML}{80CCCC}
\providecommand{\tabularnewline}{\\}

\makeatother

\begin{document}

\title{Quantum gravity inspired nonlocal gravity model}

\author{Luca Amendola}

\affiliation{Institut f\"{u}r Theoretische Physik, Ruprecht-Karls-Universit\"{a}t Heidelberg,
Philosophenweg 16, 69120 Heidelberg, Germany}

\author{Nicol\`{o} Burzill\`{a}}

\affiliation{Institut f\"{u}r Theoretische Physik, Ruprecht-Karls-Universit\"{a}t Heidelberg,
Philosophenweg 16, 69120 Heidelberg, Germany}

\affiliation{Dipartimento di Fisica e Astronomia, Universit\`{a} degli Studi di Catania, Via S. Sofia 64, 95123 Catania, Italy}




\author{Henrik Nersisyan}

\email{h.nersisyan@thphys.uni-heidelberg.de}

\affiliation{Institut f\"{u}r Theoretische Physik, Ruprecht-Karls-Universit\"{a}t Heidelberg,
Philosophenweg 16, 69120 Heidelberg, Germany}
\begin{abstract}
We  consider a nonlocal gravity model motivated by nonperturbative quantum gravity studies. This model, if correct, suggests the existence of strong IR relevant effects which can lead to an interesting late time cosmology. We implement the IR modification directly in the effective action. We show that, upon some assumptions on initial conditions,  this model  describes an observationally viable background  cosmology being also consistent with local gravity tests.
\end{abstract}

\keywords{modified gravity, nonlocal gravity, dark energy, background cosmology}

\maketitle

\section{Introduction}

After the discovery~\cite{Riess:1998cb,Perlmutter:1998np,Sherwin:2011gv,Dunkley:2008ie,Komatsu:2008hk,vanEngelen:2012va,Scranton:2003in,Sanchez:2012sg}
of the late time acceleration of the universe, many suggestions
how to explain it have been advanced. As is well known, the most
elegant and simple solution is to include the so-called cosmological
constant $\Lambda$ into the Einstein-Hilbert action. Although this
model works extremely well at the classical level, it faces dramatic
challenges once we make a step toward quantum physics. Indeed, even
in the quasiclassical approximation where gravity is still classical
and we quantize just the matter sector, the cosmological constant
receives quantum radiative corrections of the order of $m^{4}$ for each massive species, which
give a big contribution to the tiny value of $\Lambda$ needed for
an appropriate cosmic evolution. In other words, $\Lambda$ is not
a technically natural parameter. 

Of course, if the theory were renormalizable, one could adjust the value
of $\Lambda$ to fit the observations and then not worry anymore about
the questions of technical naturalness. Unfortunately, the Einstein-Hilbert
theory with a cosmological constant is not perturbatively renormalizable, and at each level of loop corrections it will receive
uncompensated quantum contributions which will destroy the predictability
of the model. In this respect, quantum gravity could solve this problem
by providing a technically natural way of cosmological constant generation.
Again, in the context of quantum gravity there could be strong IR-relevant effects which can lead to an
interesting modification of the standard cosmological constant scenario and
provide a dynamical dark energy candidate \cite{Wetterich:2017ixo}. In this direction, in Refs.~\cite{Hamber:2005dw,Hamber:2006sv},
the authors have argued that nonperturbative lattice quantum gravity calculations
can lead to a situation where the gravitational interactions slowly
increase with distance. This behavior is encoded in the running of the
gravitational constant $G$. The running of $G$ is calculated in momentum
space and in order to write a corresponding running in coordinate
space one has to specify what is the relevant cutoff. 

The choice of the relevant cutoff is not unique and in principle if
we want to have a general covariance at the level of the effective
action, we can choose as a cutoff an arbitrary covariant function
which scales as $k^{2}$. In literature, common choices have been
either $k^{2}\sim R$ (see e.g. Refs.~\cite{Bonanno:2017pkg,Bonanno:2015fga}) or $k^{2}\sim\Box$. In this paper we explore
the choice $k^{2}\sim\Box$, a choice that renders the effective action nonlocal. Another important point is the inclusion
of a nonperturbative scale $\zeta$, the scaling of which can be
approximated with an inverse Hubble function $H^{-1}$. The appearance
of this scale makes the model to be relevant for IR-dynamics of our
universe and can lead to an interesting phenomenology for the late-time
universe. Previously, in Refs.~\cite{Hamber:2005dw,Hamber:2006sv},
the authors have studied the late time cosmology of this model by
replacing Newton's constant with
a running one at the level of Friedman equations. In this work, we study this model in a more
self-consistent way, which is to get the equations of motions (EoMs)
by varying an effective action. In this case, one has to vary the
d' Alembert operator, which gives rise to additional terms in Friedman's
equations. For a particular choice of a critical exponent $\nu$ this
model predicts  results for the late time cosmology very similar to
those of Maggiore and Mancarella's ($RR$) nonlocal gravity model
\cite{Maggiore:2014sia}.

Throughout the paper, we work in flat space and natural units, i.e.
units such that $c=\hbar=1$. Furthermore, we will denote with a ``dot'' derivative with
respect to the cosmic time and with a ``prime'' derivative with respect to the number of e-foldings.

\section{The Model}

\noindent The structure of the model derives from the studies of nonperturbative lattice
quantum gravity. Ref. \cite{Hamber:2005dw,Hamber:2006sv,Hamber:1994jh} have argued that the RG improvement
of the gravitational constant $G$ leads to the following effective
action in coordinate space\footnote{The correct effective action should also contain the running cosmological constant term $\Lambda_{k}$. As is argued in Refs.~\cite{Hamber:2004ew,Hamber:2005dw} in pure lattice gravity the bare cosmological constant $\Lambda$ is scaled out and does not run. So, in this case $\Lambda$ is a constant which has to be properly fixed. Similarly to studies in Ref.~\cite{Hamber:2005dw}, in this work will assume that the contribution from $\Lambda$ is  subdominant with respect to the one induced by the running of $G$, so that we can effectively approximate $\Lambda \approx 0$.}
\begin{equation}
\frac{1}{16\pi G}\int d^{4}x\sqrt{-g}\left(1-c_{\zeta}(\frac{1}{\zeta^{2}\Box})^{1/2\nu}+O((\zeta^{2}\Box)^{-1/\nu})\right)R,\label{eq:runningG}
\end{equation}
where the relevant cutoff is provided in the form $k^{2}\sim\Box$.
In the action~(\ref{eq:runningG}) the constant $\nu$ stands for
a critical exponent, which is defined as 
\begin{equation}
\nu=-(\beta'(G_{\text{c}}))^{-1},
\end{equation}
with the $\beta$ function calculated in the vicinity of the UV non-Gaussian
fixed point (NGF) $G_{\text{c}}$. The critical exponent $\nu$ in general
is a positive rational number and highly depends on the scheme of
calculation. In realistic scenarios the value of $\nu^{-1}$ belongs
to the interval $\nu^{-1}\in\left[1,4\right]$~\cite{Litim:2003vp,Brezin:1976ap,Reuter:2001ag}. Another parameter to be specified in the action~(\ref{eq:runningG})
is the genuinely nonperturbative scale $\zeta$. Indeed, $\zeta$
is defined as 
\begin{equation}\label{zeta}
\zeta^{-1}\approx\Lambda_{\text{cut}}\exp(-\int^{G}\beta(G')^{-1}dG')\sim_{G\longrightarrow G_{\text{c}}}\Lambda_{\text{cut}}|G-G_{\text{c}}|^{\nu},
\end{equation}
where $\Lambda_{\text{cut}}$ is the UV cutoff of the theory\footnote{For the case of lattice quantum gravity the UV cutoff corresponds to the inverse lattice spacing, i.e. $\Lambda_{\text{cut}}\sim l_{\text{p}}^{-1}$.}. To determine the
real physical value of $\zeta$ we need to have some physical input
since the underlying theory  cannot fix it. An important property of $\zeta$ defined in Eq.~(\ref{zeta}) is  that when we move away from the UV fixed point along a RG trajectory, and in the case of positive critical exponents $\nu$, we have that $\zeta^{-1}$ grows. This behavior tells us that the corresponding QG corrections to the Einstein-Hilbert action, which are proportional to $\zeta^{-1}$, might enter into the strongly coupled phase and become dominant at late times when we run towards IR scales along RG trajectories. Now, in order to associate the scale of $\zeta$ with a scale of some physical quantity we will follow the discussion in~\cite{Hamber:2006sv} where it has been argued that at the late-time cosmological setup it is natural to associate this nonperturbative scale with either
the inverse of the average curvature $\left\langle R\right\rangle $
or with the inverse of the Hubble function, which determines the macroscopic
size of the universe. As in previous works \cite{Hamber:2005dw,Hamber:2006sv}, here
we will select the second option, so that $\zeta\approx H_{0}^{-1}$, where $H_{0}$ is the current value of the Hubble parameter. The only parameter left
in the action~(\ref{eq:runningG}) is then $c_{\zeta}$. The value of
$c_{\zeta}$ can be estimated in a lattice quantum gravity theory
and, in contrast to $\nu$, which has a universal value, it depends on the choice of the regularization scheme
and in general is estimated to be a order one parameter. So the only
free parameter of the model at hand is the nonperturbative scale $\zeta$,
which has to be fixed in such a way to provide a valid cosmology. 

Depending on the value of $\nu$ we can have either rational or integer
powers of $(1/\Box)$ operator in the action. In general it is not
trivial how to deal with a rational power of a differential operator, although there are some works elaborating on this issue (see, e.g. Refs.~\cite{Barnaby:2007ve, LopezNacir:2006tn,Hamber:2005dw}). In this
work, however, we will study only the cases when the power of the $\Box$
operator is an integer number, leaving the treatment of the general rational power of the d'Alembert operator to  future works. In the range  $\nu^{-1}\in\left[1,4\right]$, an integer value for $1/(2\nu )$
can be realized either when $\nu^{-1}=2$ or $\nu^{-1}=4$. A model
similar to the case with $\nu^{-1}=2$ has been recently studied in
Ref.~\cite{Vardanyan:2017kal}. So, in this work we will mainly
concentrate on the case when $\nu^{-1}=4$\footnote{Lattice quantum gravity calculations suggest for the value of the critical exponent  $\nu^{-1}=3$. In this respect our choice of $\nu^{-1}=4$ might be justified as follows. Reference~\cite{Vardanyan:2017kal} shows that the model with $\nu^{-1}=2$ leads to a valid cosmology. As will be shin later,  this outcome holds also for the current model with $\nu^{-1}=4$. Hence, one can conclude that if the model has a valid cosmology for the endpoints of the interval $\nu^{-1}\in\left[2,4\right]$, which includes also the case $\nu^{-1}=3$, it is probably valid also for other points of the same interval. Of course, this outcome can only hold if the model does not exhibit instabilities for particular values of $\nu$ within that interval. In any case, further detailed studies are needed in order to get a clear picture about the cosmology of the model with $\nu^{-1}=3$. }. For this choice of $\nu$
the effective action~(\ref{eq:runningG}) reduces to
\begin{equation}
S=\frac{1}{16\pi G}\int d^{4}x\sqrt{-g}\left(1-\frac{M^{4}}{6}\frac{1}{\Box^{2}}\right)R+\int d^{4}x\sqrt{-g}\mathcal{L}_{m},\label{eq:zero_one_action2}
\end{equation}
where we have also added the general matter Lagrangian. As already
mentioned, the cosmological implementation of this type of model has
already been studied in \cite{Hamber:2005dw,Hamber:2006sv}. In these
works, for simplicity, the running of the gravitational constant was
directly embedded into the right-hand side of the Friedman equations.
Below, we will show that  doing so
one  loses terms in equations of motion which can have a significant
impact on the cosmological evolution. Another important issue which arises when we directly implement the running of the gravitational constant into the Friedman equations, is related to a violation of the covariant conservation of the energy-momentum tensor. Indeed, if the energy-momentum tensor of the matter sector is derived from a covariant action, it will be automatically  covariantly conserved. On the other hand, from the Bianchi identities, we know that the Einstein tensor is also covariantly conserved. So,  if  in the Friedman equations we change the gravitational constant with a running one obtained through  an inverse-box structure,  we will not have anymore a covariant conservation on the matter side of the  equations, as the covariant derivative $\nabla_{\mu}$  and the inverse d'Alembert operator $\Box^{-1}$ do not commute in general, $\left[\nabla_{\mu},\Box^{-1}\right]\neq0$.

\section{The $\Box^{-2}R$ model}

\noindent To study the cosmological evolution of this model we need
to derive the Friedman equations. In Appendix~\ref{appendix 1}
we have derived the EoMs for a more general model. Corresponding EoMs
for the $\Box^{-2}R$ model can be deduced by inserting $p=0$ and
$n=1$ into Eqs.~(\ref{eq:nleinstein}-\ref{auxgen}), which
will give us

\begin{equation}
\begin{split}G_{\alpha\beta}=\frac{M^{4}}{6} & \left\lbrace \vphantom{\frac{1}{1}}LR_{\alpha\beta}-\nabla_{\alpha}\nabla_{\beta}L-g_{\alpha\beta}Q-\frac{1}{2}g_{\alpha\beta}[S+RL]+\right.\\
 & \left.+\frac{1}{2}g_{\alpha\beta}g^{\sigma\lambda}[\nabla_{\sigma}Q\nabla_{\lambda}S+\nabla_{\sigma}U\nabla_{\lambda}L]+\right.\\
 & \left.-\nabla_{\alpha}U\nabla_{\beta}L-\nabla_{\beta}Q\nabla_{\alpha}S-\frac{1}{2}g_{\alpha\beta}UQ\right\rbrace +8\pi GT_{\alpha\beta},
\end{split}
\label{eq:fg_nleinstein2}
\end{equation}
with the four auxiliary fields satisfying the following set of equations
\begin{align}
\Box U & =-R, & \Box Q & =-1,\nonumber \\
\Box S & =-U, & \Box L & =-Q.\label{eq:aux_fields1}
\end{align}
In Eq.~(\ref{eq:fg_nleinstein2}), $G_{\alpha\beta}$ stands for the
Einstein tensor and $T_{\alpha\beta}$ is a perfect fluid energy-momentum
tensor defined as $T_{\alpha}^{\beta}={\rm diag}(-\rho,p,p,p)$, where $\rho$
and $p$ are correspondingly the energy density and pressure of the
fluid.

In the case of the $\Box^{-2}R$ model the auxiliary fields $Q$ and
$L$ have a simple meaning, namely, they are the Lagrange multipliers
of the constraint equations. Indeed, let us write the gravitational
part of our nonlocal action (\ref{eq:zero_one_action2}) in a local
way by introducing the constraint equations right at the level of
the action. Then we will have for the gravitational part 
\begin{equation}
S=\frac{1}{16\pi G}\int d^{4}x\sqrt{-g}\left[R-\frac{M^{4}}{6}S+\alpha_{1}(\Box U+R)+\alpha_{2}(\Box S+U)\right],\label{eq:zero_one_local}
\end{equation}
where $\alpha_{1}$ and $\alpha_{2}$ are the Lagrange multipliers.
Now by taking the variation of the action (\ref{eq:zero_one_local})
with respect to the Lagrange multipliers and the auxiliary fields
$U$ and $S$, we get the following set of equations 
\begin{align}
\Box U & =-R, & \Box\alpha_{1} & =-\alpha_{2},\nonumber \\
\Box S & =-U, & \Box\alpha_{2} & =\frac{M^{4}}{6}.\label{eq:aux_localfield}
\end{align}
From Eq.~(\ref{eq:aux_localfield}) we see that $\alpha_{1}$ and
$\alpha_{2}$, up to the multiplicative constant $M^{4}/6$, correspond
to $L$ and $Q$, respectively. Defining the dimensionless fields

\begin{equation}
V=H_{0}^{2}S\qquad W=H_{0}^{2}Q\qquad Z=H_{0}^{4}L,
\end{equation}
and assuming a flat FLRW metric, the $(00)$ component of Eq.~(\ref{eq:fg_nleinstein2})
becomes 
\begin{equation}
h^{2}=\frac{\gamma}{4}\left\lbrace V+WU+h^{2}[6Z+6Z'-U'Z'-V'W']\right\rbrace +\Omega_{\text{R}}^{(0)}e^{-4N}+\Omega_{\text{M}}^{(0)}e^{-3N},\label{eq:fg_nleinstein2_zero}
\end{equation}
where $\gamma=(1/9)(M/H_{0})^{4}$ and $h=H/H_{0}$.

In Eq.~(\ref{eq:fg_nleinstein2_zero}), $\Omega_{\text{R}}^{(0)}$ and $\Omega_{\text{M}}^{(0)}$
are the current values of radiation and matter critical densities
in the universe, respectively. By using the identity 
\begin{equation}
\Omega=\Omega_{\text{R}}^{(0)}e^{-4N}+\Omega_{\text{M}}^{(0)}e^{-3N}=h^{2}(\Omega_{\text{R}}+\Omega_{\text{M}}),
\end{equation}
from Eq.~(\ref{eq:fg_nleinstein2_zero}) we finally get for $h^{2}$
\begin{equation}
h^{2}=\frac{(\gamma/4)(V+WU)}{1-\Omega_{\text{M}}-\Omega_{\text{R}}-(\gamma/2)[3Z+3Z'-(1/2)(U'Z'+V'W')]}.\label{eq:fg_nleinstein2_zero3}
\end{equation}
The set of differential equations (\ref{eq:aux_fields1}) for the
auxiliary fields, assuming homogeneity, are now written as 
\begin{align}
 & U''=-(3+\xi)U'+6(2+\xi),\label{eq:u_pprime}\\
 & V''=-(3+\xi)V'+\frac{U}{h^{2}},\label{eq:v_pprime}\\
 & W''=-(3+\xi)W'+\frac{1}{h^{2}},\label{eq:w_pprime}\\
 & Z''=-(3+\xi)Z'+\frac{W}{h^{2}}.\label{eq:z_pprime}
\end{align}
where $\xi$, defined as $\xi=h'/h$, has the following structure:
\begin{equation}
\xi=\frac{1}{2(1-(3/2)\gamma Z)}\left[\frac{\Omega'}{h^{2}}+\frac{3}{2}\gamma\left(\frac{W}{h^{2}}-4Z'+U'Z'+V'W'\right)\right].\label{eq:xi_expr}
\end{equation}

\section{numerical solutions}

\noindent Before moving to the full numerical analysis of the system,
here we can briefly comment upon the stability of the system during
matter and radiation domination periods. In this case, $\xi$ can
be well approximated with a constant in each era: $\xi_{0}=\left\lbrace -2,-3/2\right\rbrace $
in matter and radiation-dominated periods, respectively. First, we
can check the consistency of the homogenous solutions of Eqs.~(\ref{eq:u_pprime}-\ref{eq:z_pprime}).
The homogenous solutions are the followings 
\begin{eqnarray}
U=u_{0}+u_{1}e^{-(3+\xi_{0})N},\\
\bar{V}=\bar{v}_{1}e^{-(3-\xi_{0})N}+\bar{v}_{2}e^{2\xi_{0}N},\\
\bar{W}=\bar{w}_{1}e^{-(3-\xi_{0})N}+\bar{w}_{2}e^{2\xi_{0}N},\\
\bar{Z}=\bar{z}_{1}e^{-(3-\xi_{0})N}+\bar{z}_{2}e^{2\xi_{0}N},
\end{eqnarray}
where $\bar{V}=h^{2}V$, $\bar{W}=h^{2}W$ and $\bar{Z}=h^{2}Z$.
From the equations above we see that when $\xi_{0}$ is changing in
the interval $\xi_{0}\in\left[-2,0\right]$ we have that the solutions
either remain constant or decrease exponentially. This ensures that
the solutions are stable during matter and radiation-dominated periods.



 To solve Eqs.~(\ref{eq:u_pprime}-\ref{eq:z_pprime}) numerically, with
the constrains~(\ref{eq:fg_nleinstein2_zero3}) and~ (\ref{eq:xi_expr}), we first fix the present values
\begin{equation}
\Omega_{\text{M}}^{(0)}=0.3\qquad\Omega_{\text{R}}^{(0)}=4.15\times10^{-5}h^{-2}
\end{equation}
to the standard values. Although the constraint on $\Omega_{\text{M}}$ has been obtained assuming standard $\Lambda$CDM and therefore in principle should be estimated anew with the present model,  we will see the background evolution  turns out not to be very different from the standard one, so our choice may be considered a reasonable approximation.  For later use, we also need to define
the effective equation of state $w_{\text{eff}}$ and the critical dark
energy density $\Omega_{\text{DE}}$, respectively, as 
\begin{equation}
w_{\text{eff}}=-1-\frac{2}{3}\xi\label{weff},
\end{equation}
\begin{equation}
\Omega_{\text{DE}}=\frac{\gamma}{4}\left[\frac{1}{h^{2}}(V+WU)+6Z+6Z'-U'Z'-V'W'\right].\label{omegaDE}
\end{equation}
Using the definition of $\Omega_{\text{DE}}$ we can rewrite Eq.~(\ref{eq:fg_nleinstein2_zero})
as 
\begin{equation}
\Omega_{\text{DE}}=1-\Omega_{\text{M}}-\Omega_{\text{R}}.
\end{equation}
Furthermore, from the continuity equation of the dark energy critical density $\Omega_{\text{DE}}$
\begin{equation}
\Omega'_{\text{DE}}+(3+3w_{\text{DE}}+2\xi)\Omega_{\text{DE}}=0,
\end{equation}
we find for the dark energy equation of state parameter $w_{\text{DE}}$ 
\begin{equation}
w_{\text{DE}}=-1-\frac{2}{3}\xi-\frac{1}{3}\frac{\Omega'_{\text{DE}}}{\Omega_{\text{DE}}}=-\frac{3+\Omega_{\text{R}}+2\xi}{3 \Omega_{\text{DE}}}.
\end{equation}
Finally, making use of Eq.~(\ref{eq:xi_expr}), we can write $w_{\text{DE}}$
explicitly in terms of the auxiliary fields 
\begin{equation}\label{wdeanaly}
\begin{split}w_{\text{DE}}= & -\frac{3+\Omega_{\text{R}}}{3\Omega_{\text{DE}}}-\frac{2}{2-3\gamma Z}\left(1-\frac{3+\Omega_{\text{R}}}{3\Omega_{\text{DE}}}+2\frac{W}{V+UW}\right)+\\
 & -\left(\frac{3\gamma}{2-3\gamma Z}\right)\frac{1}{3\Omega_{\text{DE}}}\left[U'Z'+V'W'-4Z'+\frac{W}{V+UW}\big(U'Z'+V'W'-6Z-6Z'\big)\right].
\end{split}
\end{equation}
The value of the only dimensionless free parameter of our model, $\gamma$, should be fixed in a such way as to satisfy the condition $h(0)=1$. This produces, for instance, the values $\gamma=\left\lbrace 0.702, 0.222, 0.030,0.015 \right\rbrace$ for the following initial conditions on $W$, $W_{0}=\left\lbrace 0, 0.5, 5,10\right\rbrace$, respectively. Now, to integrate Eqs.~(\ref{eq:u_pprime}-\ref{eq:z_pprime}) we need to specify initial conditions on our auxiliary fields at the onset of integration deep inside the radiation-dominated period. We choose $N_{\text{in}}=-14$ as the initial time. As is also argued in Ref.~\cite{Nersisyan:2016hjh}, the choice of initial conditions for the auxiliary fields in a deep radiation-dominated period are \textit{per se} arbitrary. Their value highly depends on the physical content of the Universe at the epoch we start evolving the differential equations~(\ref{eq:u_pprime}-\ref{eq:z_pprime}). For simplicity, we will assume that all the auxiliary fields in our model, apart from $W$, have vanishing initial conditions. The reason behind this particular choice of initial conditions will become clear below.

\begin{figure}[hpt]
\includegraphics[width=8.4cm]{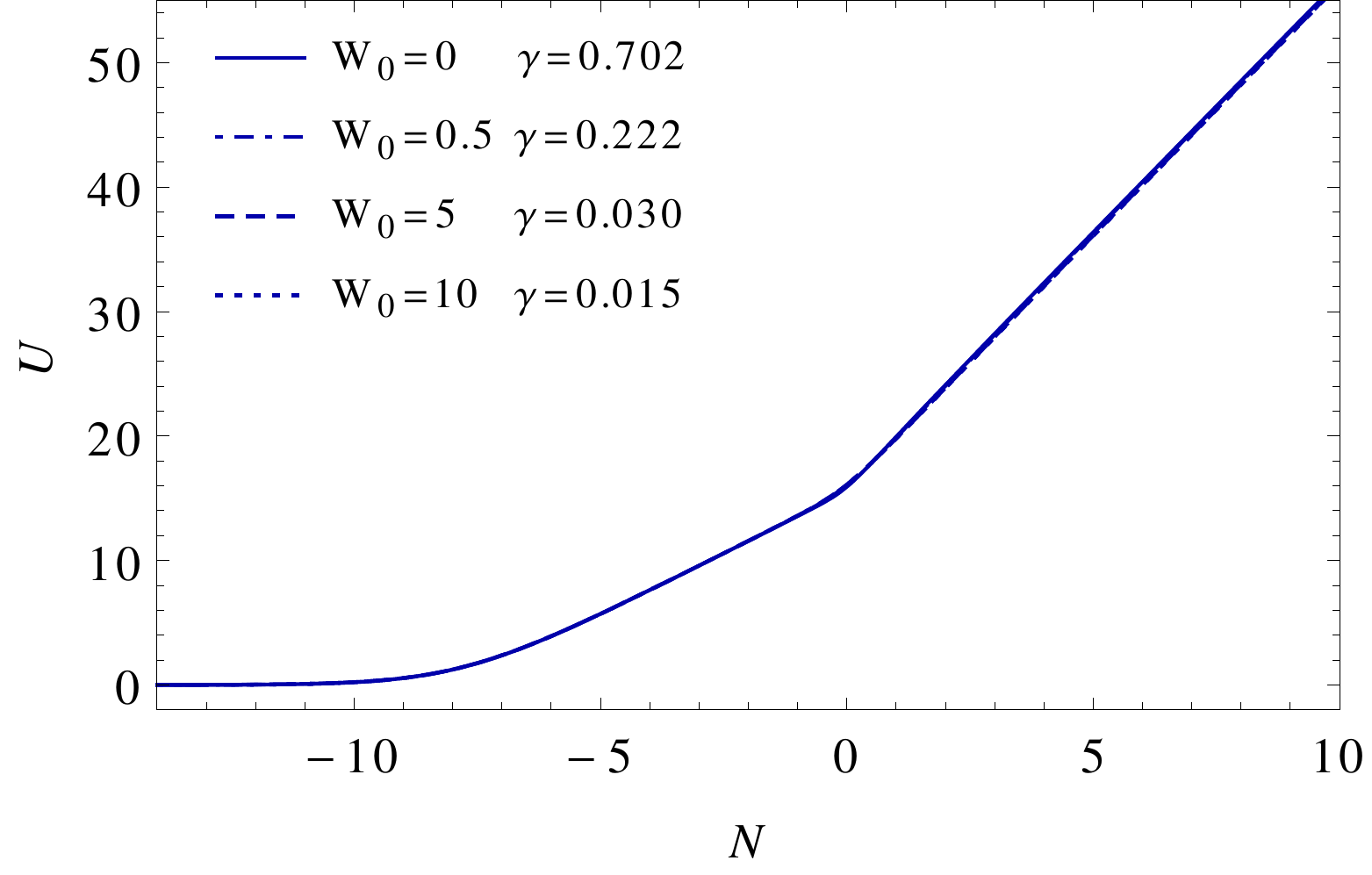}\qquad{}\includegraphics[width=8.5cm]{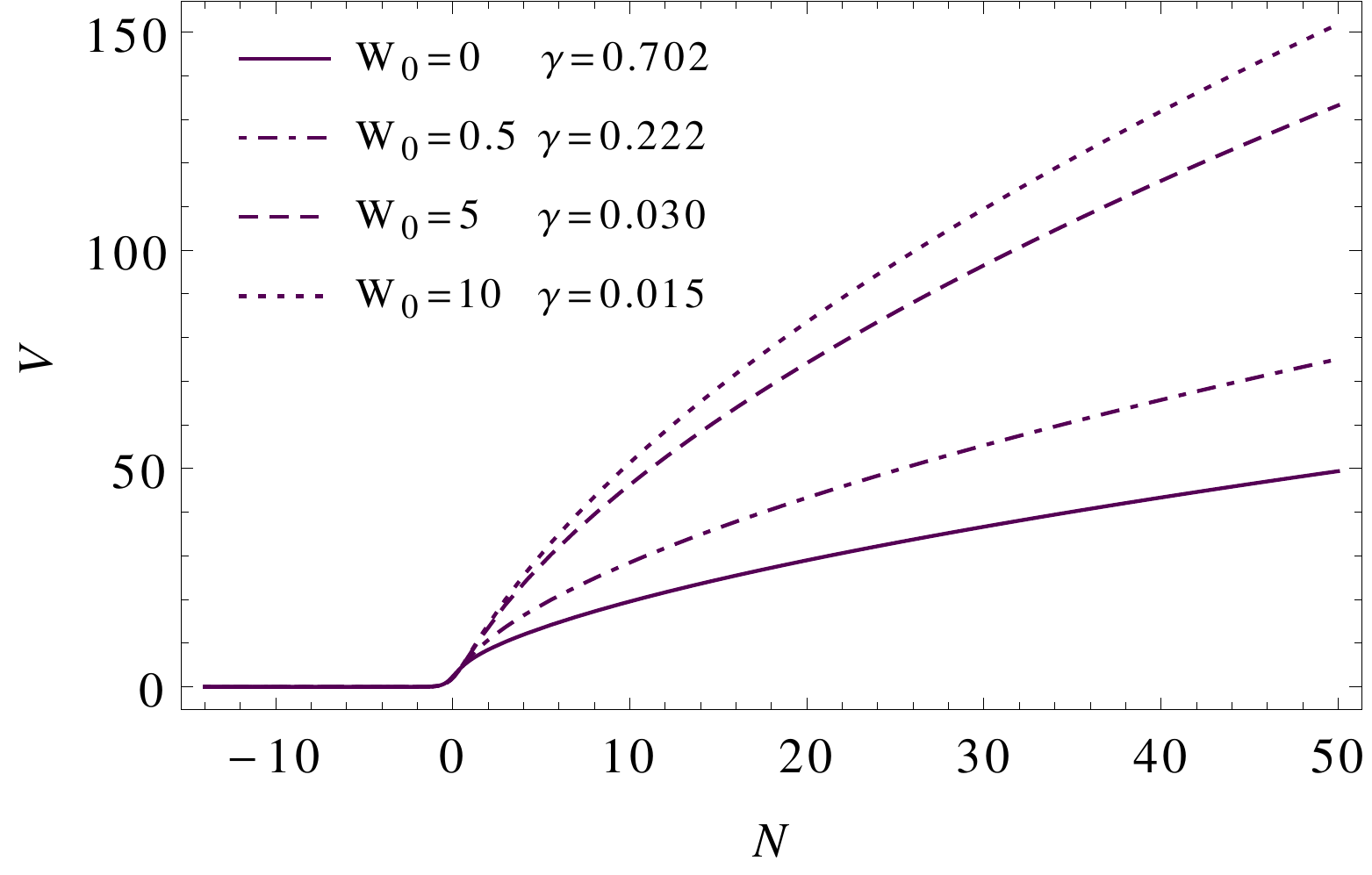}
\center \includegraphics[width=8.5cm]{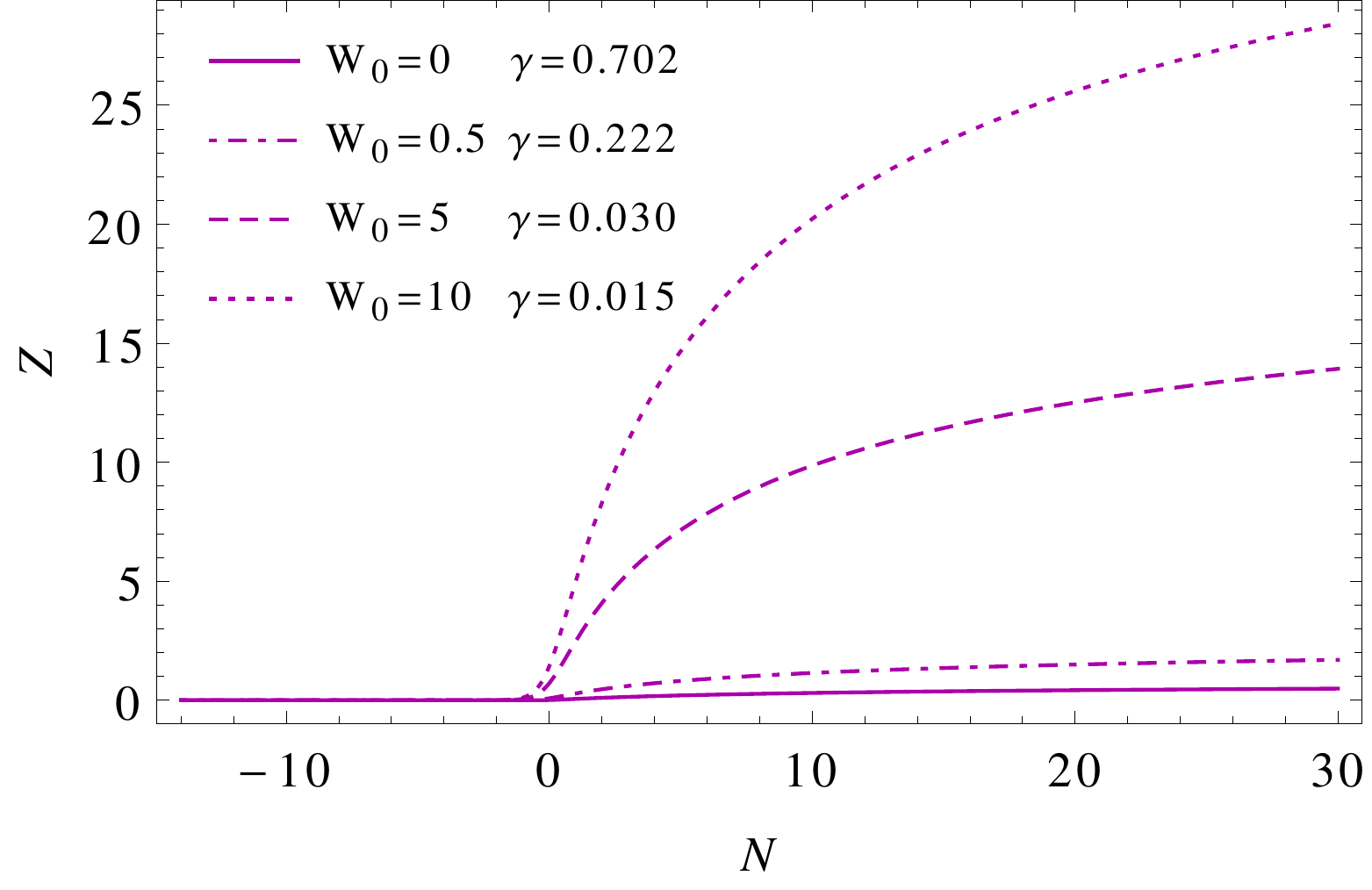}
\caption{Evolution of auxiliary fields $U$, $V$ and $Z$ as a function of $N=\ln a$, for different values of $W_{0}$ and corresponding $\gamma$.
Initial conditions: $U_{0}=0$, $V_{0}=0$, $Z_{0}=0$, $U_{0}'=0$,
$V_{0}'=0$, $W_{0}'=0$, $Z_{0}'=0$ .}

\label{UpathA} 
\end{figure}

\begin{figure}[hpt]
\includegraphics[width=8.4cm]{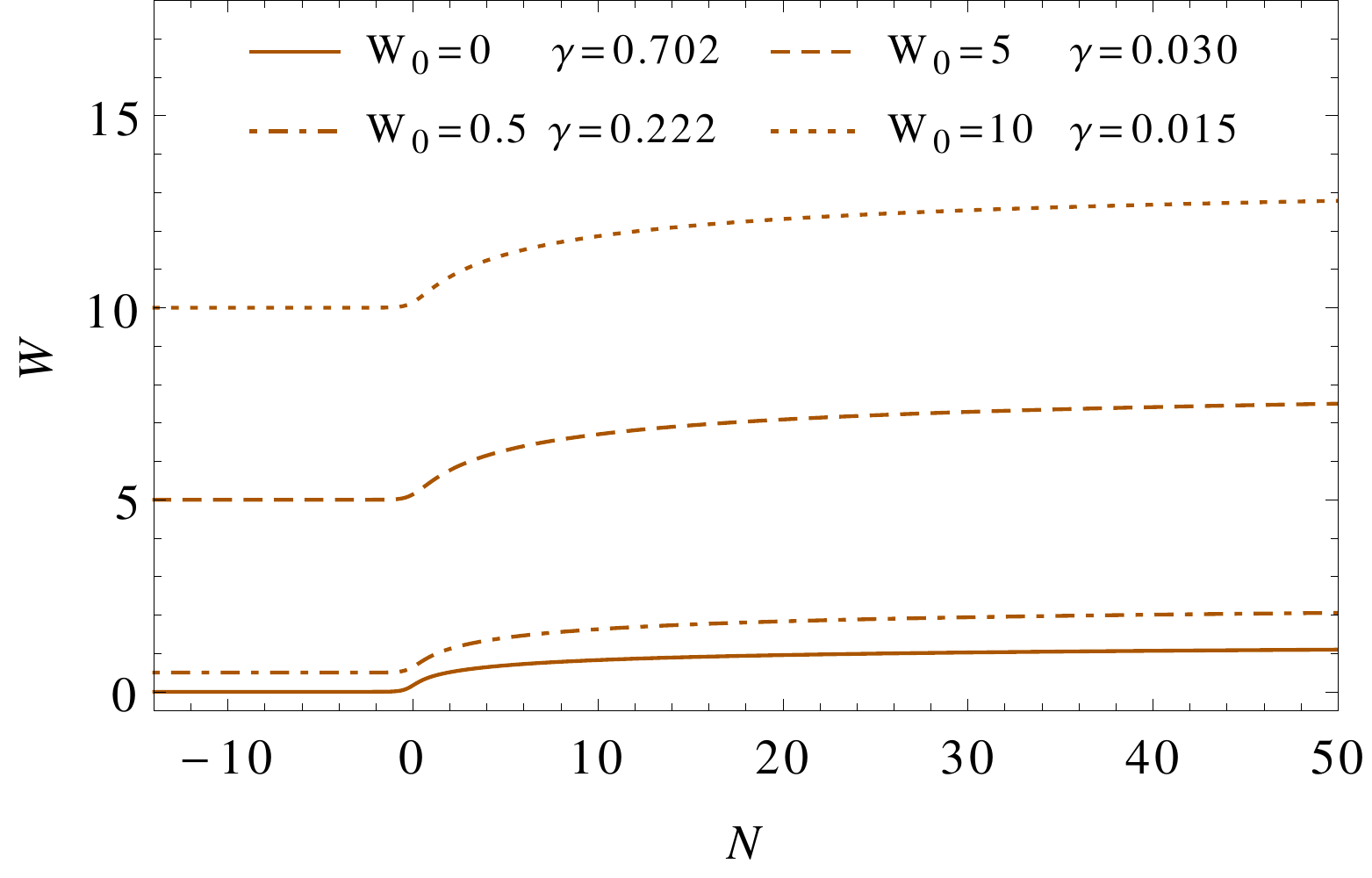}\qquad{}\includegraphics[width=8.5cm]{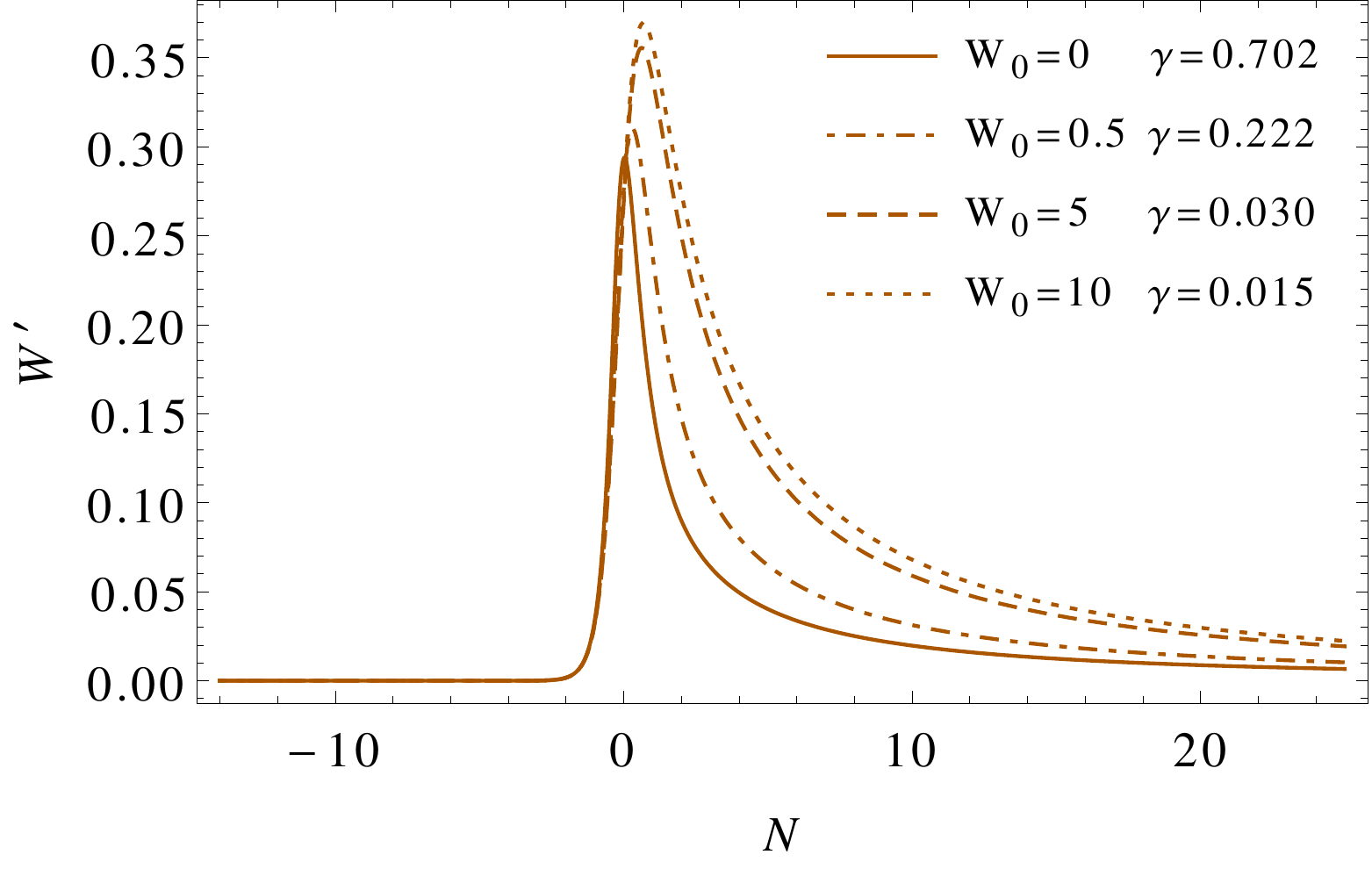}
\caption{Evolution of the auxiliary field $W$, and its first derivative $W'$ as a function of $N=\ln a$, for different values of $W_{0}$ and corresponding $\gamma$.
Initial conditions: $U_{0}=0$, $V_{0}=0$, $Z_{0}=0$, $U_{0}'=0$,
$V_{0}'=0$, $W_{0}'=0$, $Z_{0}'=0$ .}

\label{Wevol} 
\end{figure}

\begin{figure}[hpt]
\includegraphics[width=8.5cm]{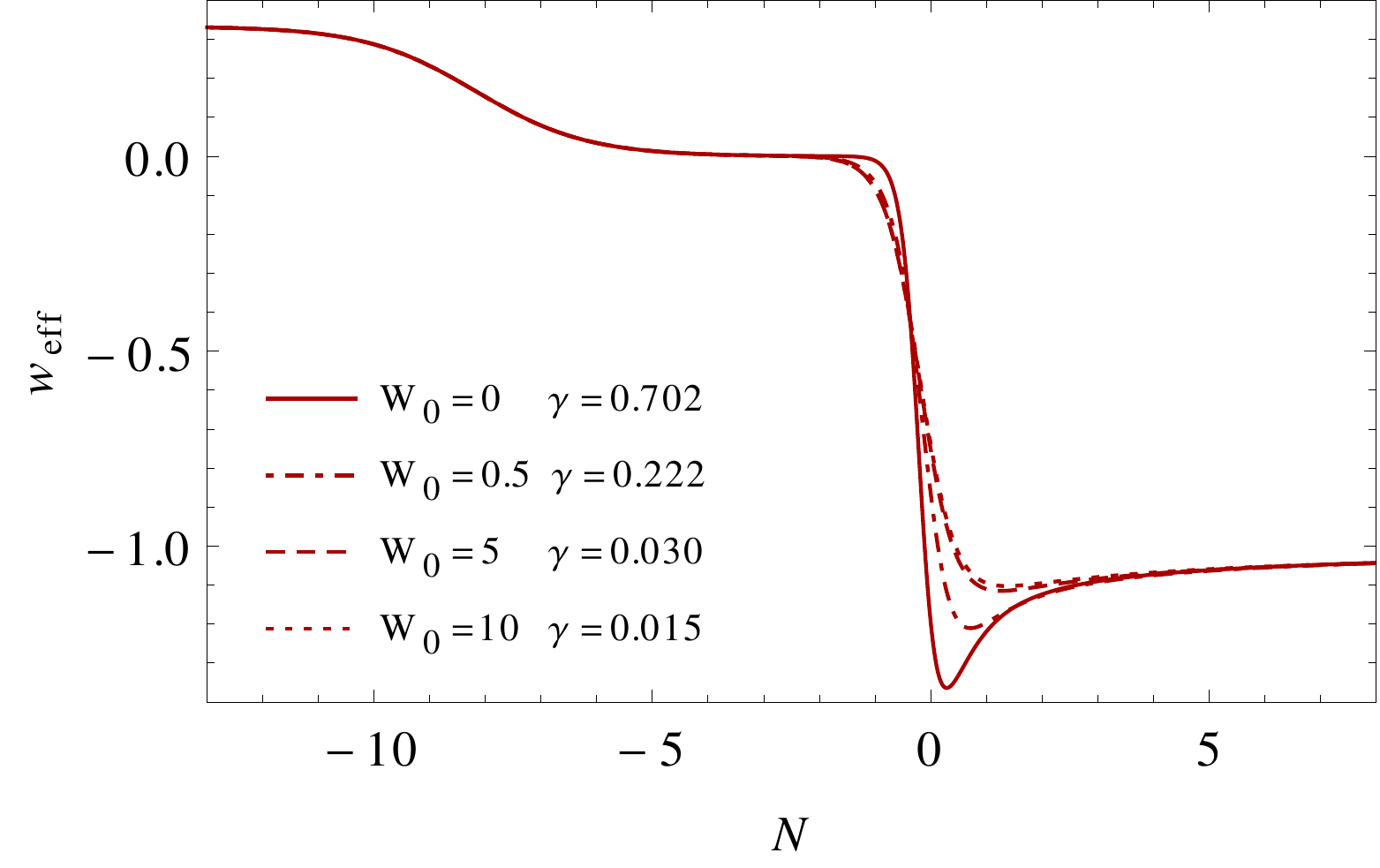}\qquad{}\includegraphics[width=8.5cm]{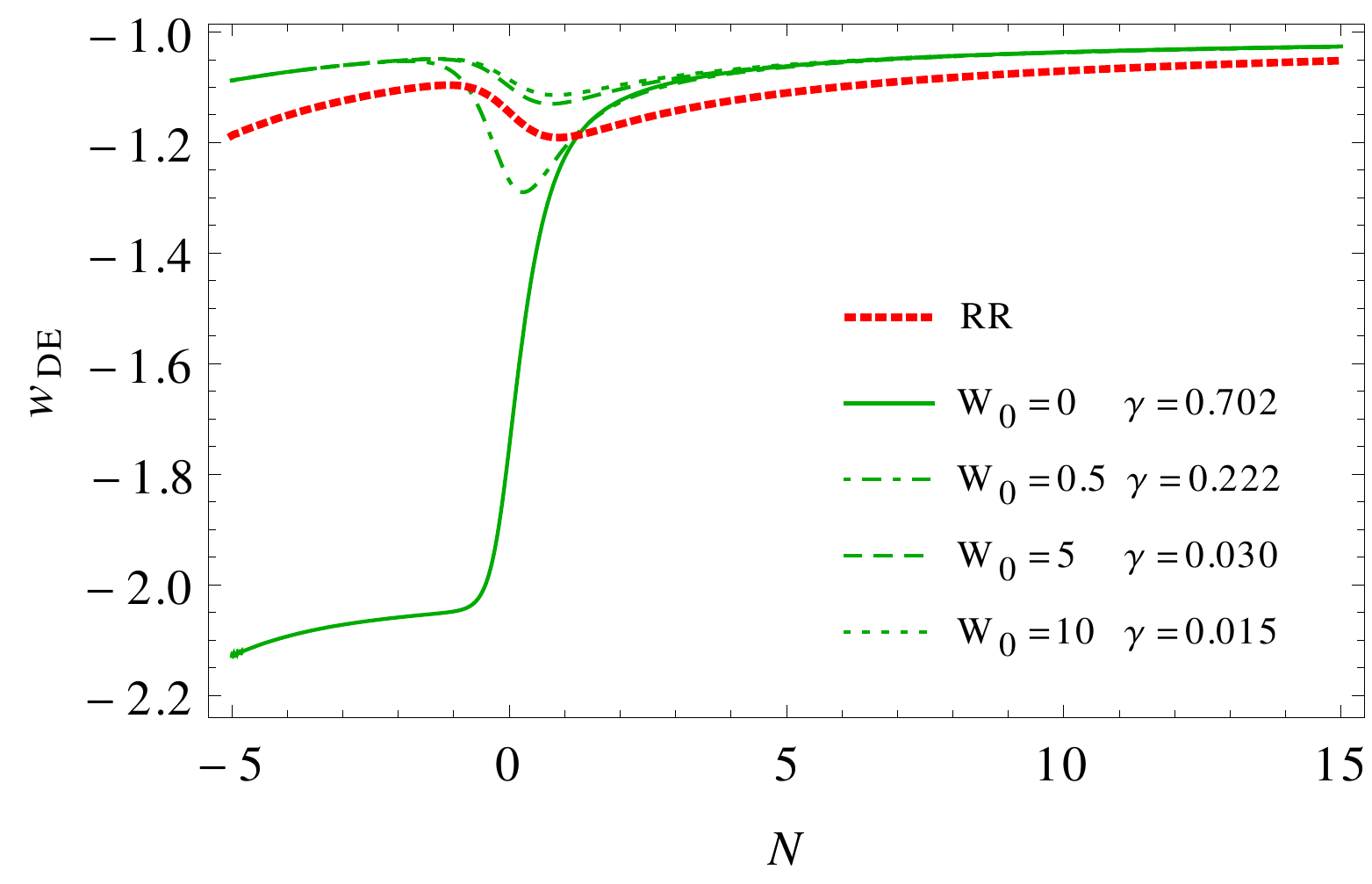}
\caption{Left panel: Evolution of $w_{\text{eff}}$  as a function of $N=\ln a$, for different values of $W_{0}$ and corresponding $\gamma$. Initial conditions:
$U_{0}=0$, $V_{0}=0$, $Z_{0}=0$, $U_{0}'=0$, $V_{0}'=0$, $W_{0}'=0$,
$Z_{0}'=0$. Right panel: Evolution of $w_{\text{DE}}$  as a function of $N=\ln a$, for different values of $W_{0}$ and the corresponding $\gamma$. Initial
conditions: $U_{0}=0$, $V_{0}=0$, $Z_{0}=0$, $U_{0}'=0$, $V_{0}'=0$,
$W_{0}'=0$, $Z_{0}'=0$.}

\label{weffcomw} 
\end{figure}

\begin{figure}[hpt]
\includegraphics[width=8.5cm]{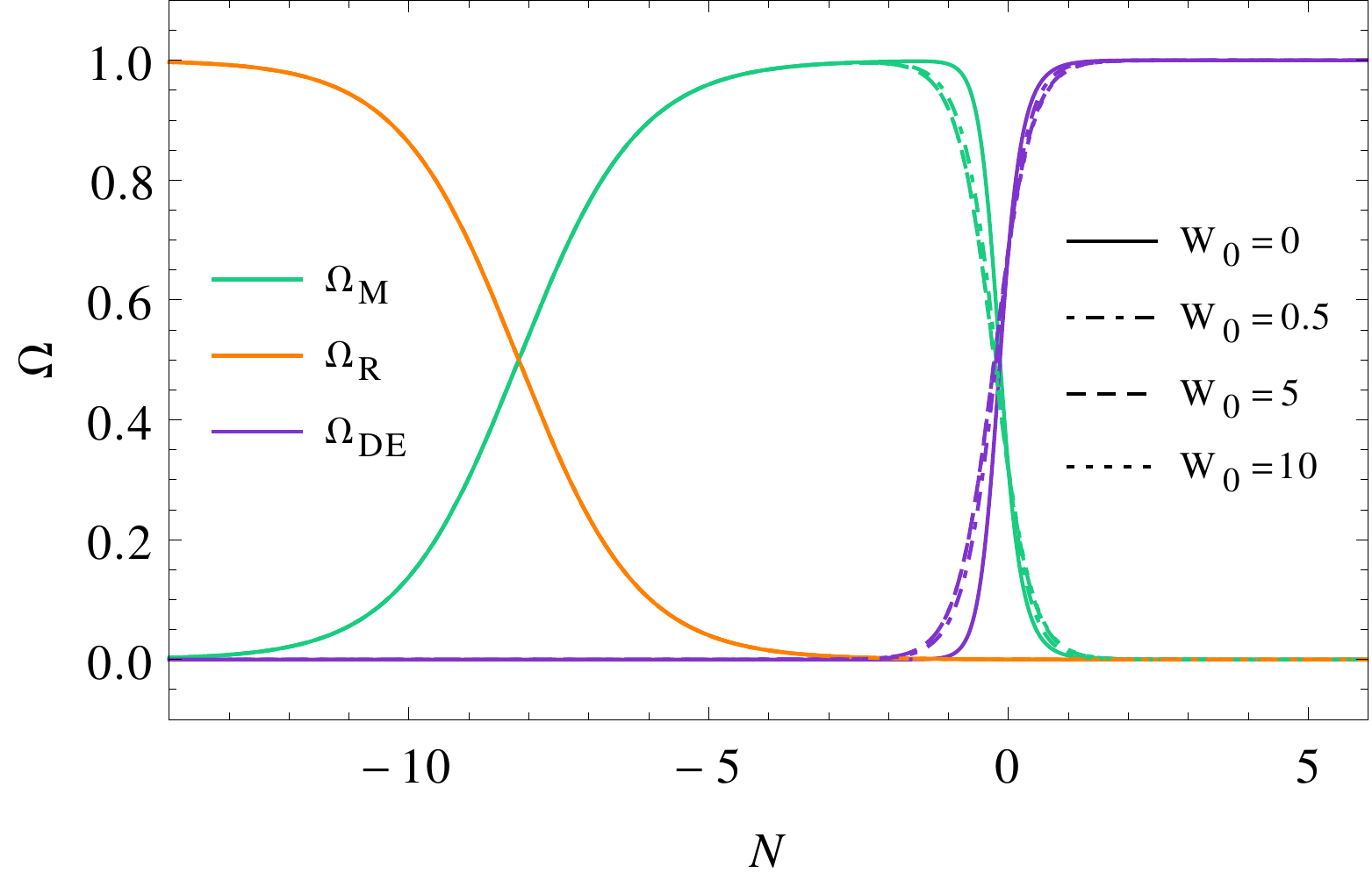}\qquad{}\includegraphics[width=8.5cm]{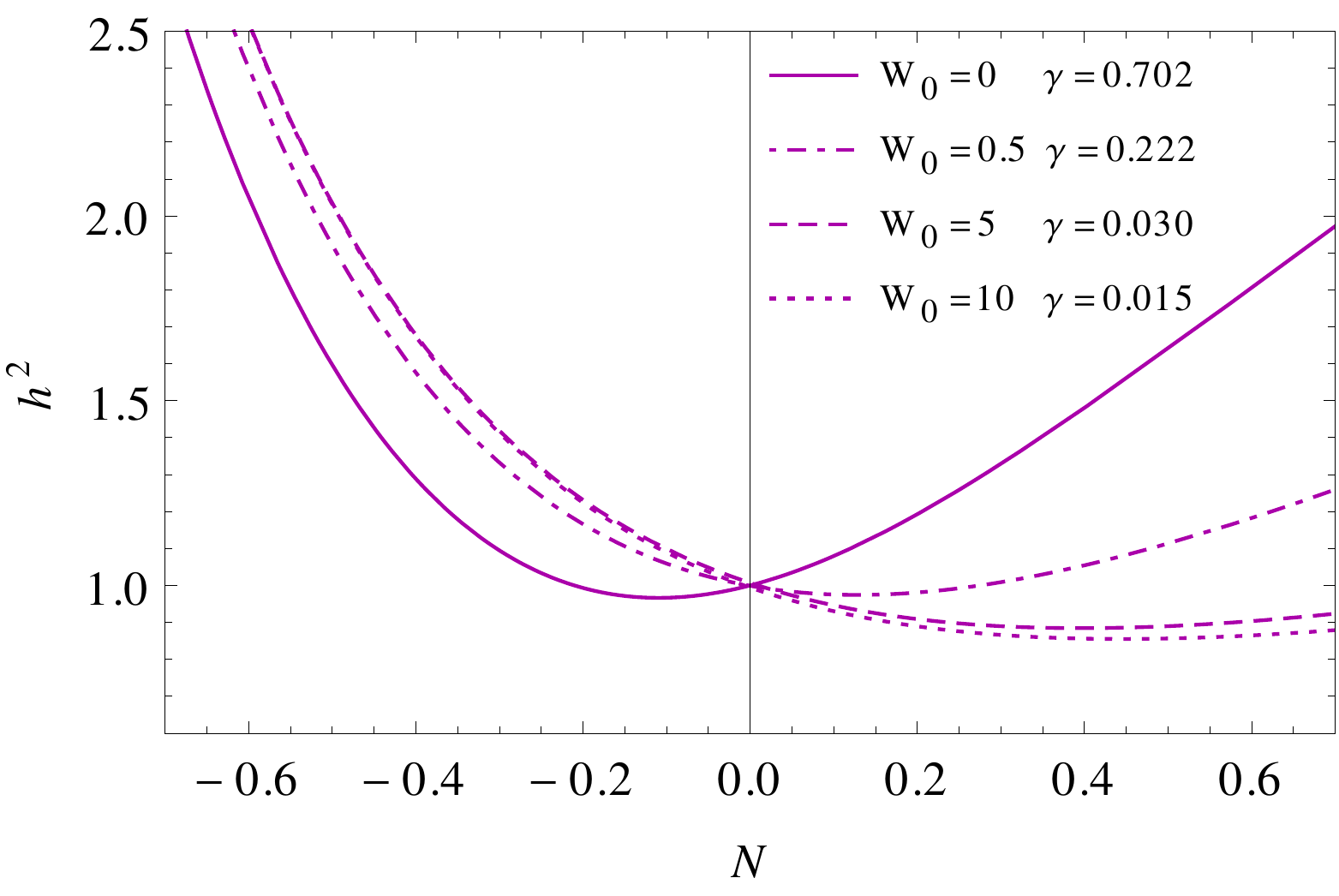}
\caption{Left panel: Evolution of $\Omega_{\text{M}}$, $\Omega_{\text{R}}$, $\Omega_{\text{DE}}$  as a function of $N=\ln a$, for different values of $W_{0}$ and corresponding $\gamma$.
Initial conditions: $U_{0}=0$, $V_{0}=0$, $Z_{0}=0$, $U_{0}'=0$,
$V_{0}'=0$, $W_{0}'=0$, $Z_{0}'=0$. Right panel: Evolution of Hubble
function $h$ .}


\label{omegah2} 
\end{figure}

Performing the integration of Eqs.~(\ref{eq:u_pprime}-\ref{eq:z_pprime}), we find the numerical results as a function of $N=\log a$ plotted in Figs.~\ref{UpathA}-\ref{omegah2} for four different choices of initial conditions for $W$. As we can see from Figs.~\ref{UpathA}-\ref{omegah2}, the evolution of all physical quantities as well as auxiliary fields does not show singularities. Moreover, from the left panel of Fig.~\ref{weffcomw} we observe that the present model predicts a well defined radiation-domination period $(w_{\text{eff}}=1/3)$, followed by a matter dominated period $(w_{\text{eff}}=0)$, which finally ends in a dark energy-dominated period. The transition  between the matter epoch and  dark energy epoch,  $(N\approx -0.4)$, is very well constrained by the current observational data~\cite{Ade:2015rim}, so one can already put some restrictions on the model. In this respect, as one can notice from Fig.~\ref{weffcomw}, in our case when the initial value of $W$~($W_{0}$) is set to be vanishing, the evolution of $w_{\text{eff}}$ and $w_{\text{DE}}$ exhibits a strongly phantom behavior. The dark energy equation of state parameter $w_{\text{DE}}$ increases sharply from  $w_{\text{DE}}\approx-2.1$ to $w_{\text{DE}}\approx-1.2$. From the observational side, constraints on $w_{\text{DE}}$ are often obtained parametrizing it as a linear function of the scale factor $a$, i.e. $w_{\text{DE}}=w_{\text{DE}}^{0}+\left(1-a\right)w_{\text{DE}}^{\text{a}}$. Comparing the values of $w_{\text{DE}}^{0}$ and $w_{\text{DE}}^{\text{a}}$ for our model from  Table~\ref{ourtable} with the corresponding observational constraints (see, e.g.,  Ref.~\cite{Suzuki:2011hu},  Table 7) we immediately see that our solution for vanishing $W_{0}$ is in strong tension with the constraints and is probably already ruled out\footnote{ Here is important to mention that these observational constraints are obtained by combining the supernovae (SNe) and cosmic microwave background (CMB) data. The constraints coming from CMB data cannot be directly applied  to our nonlocal model, both because they are based on $\Lambda$CDM and because the linear parametrization of $w_{\text{DE}}$ is not a good approximation to the evolution of the dark energy  in the past. Lifting the CMB constraints the error bars on $w_{\text{DE}}^{0}$ and $w_{\text{DE}}^{\text{a}}$, relax considerably, but even in the most realistic case the nonlocal model with vanishing $W_{0}$ will be still highly disfavored.}.

\begin{table}[t]

\centering{} %
\begin{tabular}{|c|c|c|c|c|}

\hline
$\gamma$ & $0.702$ & $0.222$ & $0.030$ & $0.015$ 
 \tabularnewline
\hline 
\hline 
$w_{\text{DE}}^{0}$ & $-1.752$ & $-1.268$ & $-1.099$ & $-1.086$ \\
\hline
$w_{\text{DE}}^{\text{a}}$ & $0.843$ & $-0.170$ & $-0.077$ & $-0.061$ \\
\hline
\end{tabular}
\caption{Today's values of $w_{\text{DE}}$ and its first derivative w.r.t. the scale factor $a$, for different values of $\gamma$ corresponding to different initial conditions on the field $W$.}
\label{ourtable}
\end{table}

As mentioned, we chose vanishing initial conditions for all auxiliary fields expect for  $W$, which means that homogeneous solutions for those auxiliary fields are set to be zero deep in the radiation domination period. By relaxing these assumptions we can see how the overall quantitative evolution is affected. This kind of analysis of initial conditions has been performed in Ref.~\cite{Nersisyan:2016hjh} in the
case of the $RR$  model. Here we apply the same analysis but only state the main outcome. We find that the behavior of the present model during the dark energy-dominated period highly depends on the choice of initial conditions for the auxiliary field $W$. Indeed, again from Fig.~\ref{weffcomw}, we observe that even a small nonvanishing value of  $W_{0}$ can efficiently soften the strongly phantom behavior of the dark energy equation of state parameter~$w_{\text{DE}}$, making it compatible with  current observational constraints.
Indeed, in contrast to the case of the  field $U$ satisfying the equation $\Box U=-R$, the field $W$ satisfies the equation $H_{0}^{2}\Box W=-1$. Therefore the  arguments in the literature ~\cite{Maggiore:2014sia} for choosing vanishing initial conditions for $U$, related to the fact that $R$ is also vanishing during the radiation-dominated period, do not hold anymore. In principle $W$ can have any initial value depending on its early history. The dependence of the evolution of the field $W$ on initial conditions is presented in the left panel of Fig.~\ref{Wevol}. 

The high sensitivity of $w_{\text{DE}}$ to the nonvanishing choice of initial conditions for $W$ can be also inferred from Eq.~(\ref{wdeanaly}). In the  expression for $w_{\text{DE}}$ we see that there are several terms which are directly proportional to  $W$. Therefore they will affect the value of $w_{\text{DE}}$ only in the cases when $W$ is nonvanishing. On the other hand, from Fig.~\ref{Wevol}, we see that  $W$ remains very close to its initial value $W_{0}$ until the matter-dark energy transition point ($N\approx -0.4$). Therefore, when $W_{0}=0$, $W$ is also vanishing, so the terms in Eq.~(\ref{wdeanaly}) proportional to $W$ will never be activated and thus will not contribute to the value of $w_{\text{DE}}$.

\section{Relation between $R\Box^{-2}R$ and $\Box^{-2}R$ models}

\noindent As one can see from Figs.~\ref{UpathA}-\ref{omegah2},
the numerical evolution of the auxiliary fields as well as the behavior
of $w_{\text{eff}}$ and $\Omega_{\text{M,R,DE}}$ are very similar to those of
the $RR$ nonlocal model presented in Refs.~\cite{Maggiore:2016gpx,Nersisyan:2016hjh}.
In order to understand why two, at first glance, completely different
models exhibit almost the same cosmological evolution let us first
compare them at the level of the actions. Here we just concentrate on the gravitational sectors:

\begin{equation}
S=\frac{1}{16\pi G}\int d^{4}x\sqrt{-g}\left[R-\frac{M^{4}}{6}\frac{1}{\Box^{2}}R\right],\label{eq:gravityaction1}
\end{equation}

\begin{equation}
S^{\text{RR}}=\frac{1}{16\pi G}\int d^{4}x\sqrt{-g}\left[R-\frac{m^{2}}{6}R\frac{1}{\Box^{2}}R\right].\label{eq:gravityaction2}
\end{equation}

By comparing the actions (\ref{eq:gravityaction1}) and (\ref{eq:gravityaction2})
we notice that when for these two models to predict the same behavior at late times one needs to have for that period

\begin{equation}\label{relmM}
m^{2}R=\beta M^{4},
\end{equation}
where $\beta$ is some constant parameter of the proportionality. The relation~(\ref{relmM}) in the language of the dimensionless parameter $\gamma$ can
be written as 
\begin{equation}
\beta^{-1} \gamma_{\text{RR}}= B \gamma,\label{gammarel}
\end{equation}
with $B \equiv H_{0}^{2}/R$. The value of the $\gamma_{\text{RR}}$
and $\gamma$ should be fixed in a way to reproduce the correct dark energy
density nowadays. For the $RR$ nonlocal gravity model the parameter $\gamma_{\text{RR}}$ is estimated to be $\gamma_{\text{RR}}\simeq 0.0089$~\cite{Maggiore:2014sia}.
\begin{figure}[tbh]
\begin{centering}
\includegraphics[width=8.5cm]{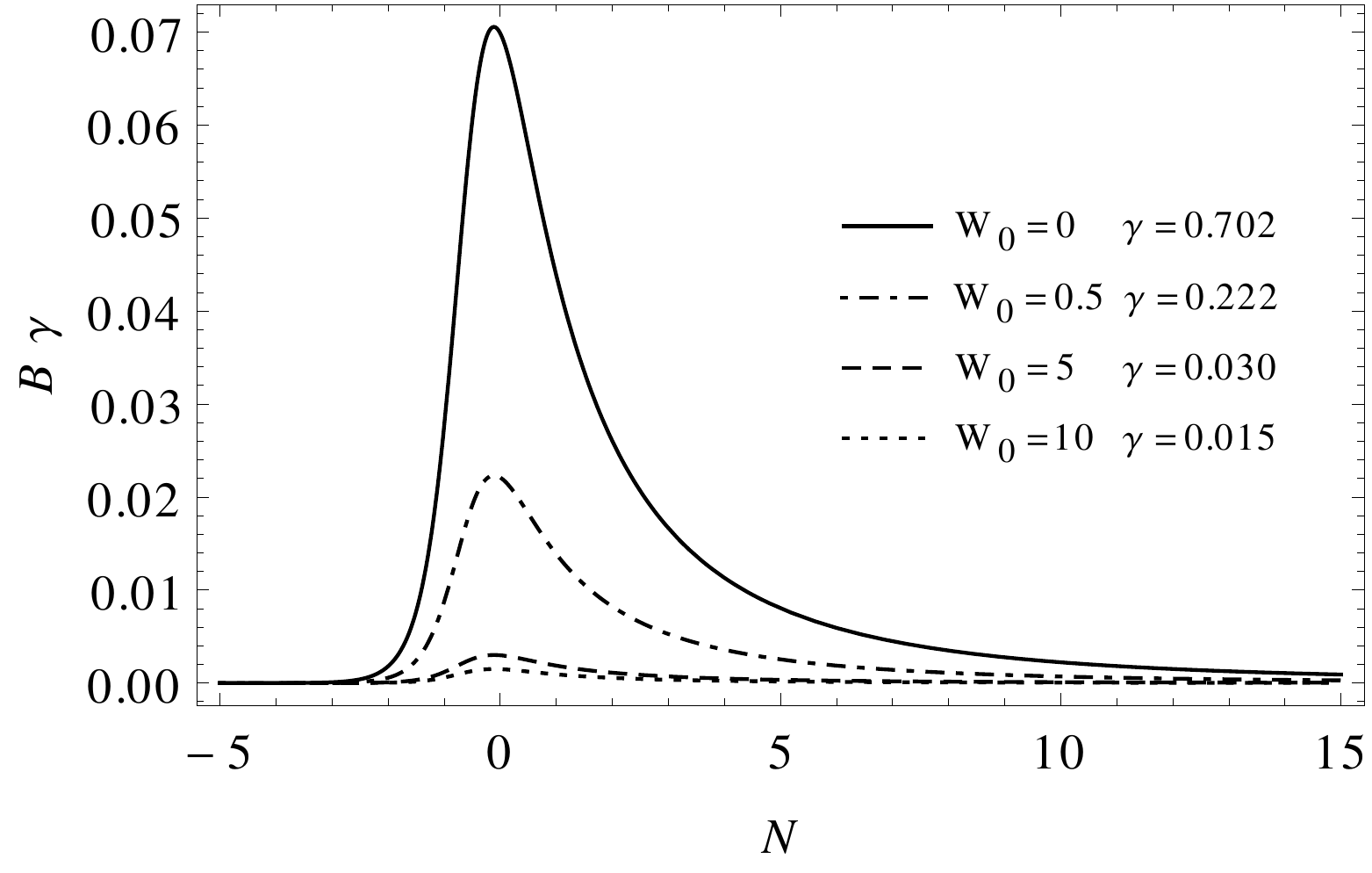}
\caption{The evolution of $B \gamma$ as a function of e-foldings
$N=\log a$ for different values of $\gamma$. }
\label{wdecomalpha} 
\end{centering}
\end{figure}
Moreover, from Fig.~\ref{wdecomalpha} we see that once
we fix the value of $\gamma$ to reproduce the exact matter content
in the universe, due to the running of $B$ the quantity $B\gamma$
grows initially and then after some time ($N\approx5$) saturates
and stays approximately constant. This means that the two theories then
should become equivalent, i.e. the condition~(\ref{gammarel}) holds. This can also be recognized from the right
panel of Fig.~\ref{weffcomw}, where we display the evolution of the dark energy
equation of state. One can see that at early times the two models
behave very differently and then around $N\approx1.5$  approach each other. 

Here an interesting question arises: if two models exhibit similar behavior at late times is there some reason to favor one model over another? In this case, to give some preference to one of the models we need to look at what mechanisms might have generated those corrections in the first place. For the present model the structure of nonlocal corrections is originated in lattice quantum gravity calculations. Moreover, calculations also show that these corrections are relevant at IR scales due to the fact that  $\zeta\approx H_{0}^{-1}$. On the other hand, for the $RR$ model presented in the Ref.~\cite{Maggiore:2014sia} there is no clear mechanism how this type of corrections can be generated from a fundamental theory. There were some suggestions that a $R\Box^{-2}R$ term can arise from the loop contributions of massless scalar fields. However, later it was shown in Ref.~\cite{Maggiore:2016fbn} that even though terms with similar structure indeed arise from perturbative calculations, due to their small coefficient they are not relevant at cosmological scales.

\section{Local gravity constraints}

Another important point to be discussed  is whether the constraints on the gravitational
constant $G$ in the solar system are satisfied. Indeed, as
it was already discussed in Ref.~\cite{Hamber:2006sv}, a vacuum-polarization-driven
running of $G$ can lead to serious difficulties with experimental
constraints on the time variability of $G$. Solar system measurements
put strong constraints on the time variation of $G$ \cite{Williams:2004qba} $\lvert\dot{G}/G\rvert<10^{-12}{\rm yr}^{-1}$.

It is important to mention that the above mentioned constraints on the time variation of the effective gravitational constant have been derived for the Earth-Moon system. In this respect it is important to know whether we can use the time variation of $G$ calculated at the cosmological scales inside the Earth-Moon system. In Refs.~\cite{Barreira:2014kra,Dirian:2016puz} it has been argued that this question should be taken with a special care. Indeed, inside the local scales such as the solar system, Earth-Moon system and etc, we do not have expansion with the Hubble rate as is the case for very large scales. This boils down to the question whether inside the solar system the scale factor $a$ in Eq.~(\ref{eq:flrwmetric}) has time dependence or not. So, if the scale factor $a$ is time dependent, the d'Alembertian operator will also depend on time so $G$ will vary inside the local scales. In the opposite case, the effective gravitational constant will be time independent so the constraints on it will be trivially satisfied. Let us also emphasize that even if on the background level, for local scales, the time dependence of $G$ can be neglected it will not guarantee that the result will be the same also on perturbative level. Indeed, possible time and coordinate dependent perturbations in the local scales can reintroduce a time dependence of $G$ which then should be consistent with all constraints. In any case, this is an open question and deserves a dedicated study which is beyond the scope of the current work.

 After simple algebraic
steps we get that for our model today's rate of $G$ is
\begin{equation}
\lvert\dot{G}/G\rvert_{0}\approx\lvert\frac{12\gamma}{2-3\gamma}\rvert H_{0}.\label{gvargam}
\end{equation}
In our case valid cosmological models are  obtained for nonvanishing values of $W_{0}$. Plugging the corresponding values of $\gamma$ into Eq.~(\ref{gvargam}) we get that indeed for all these cases $\lvert\dot{G}/G\rvert_{H_{0}}\lesssim H_{0}\lesssim 10^{-12}{\rm yr}^{-1}$.
So, the models which have a valid cosmological evolution satisfy the local constraints too.

\section{Conclusions} 

Nonlocal cosmological models have been the topic of intense work in the last few years~\cite{Dirian:2017pwp,Nersisyan:2017mgj,Nersisyan:2016jta,Maggiore:2016gpx,Foffa:2013vma,Dirian:2014ara}. They can be seen as an attempt at capturing  quantum corrections to the Einstein-Hilbert Lagrangian and to provide, at the same time, an accelerated cosmology even in the limit of vanishing cosmological constant. In this paper  we have studied in detail the background cosmological evolution of a novel nonlocal model inspired by a quantum gravity induced nonperturbative effective action in which the FRG running of the gravitational constant, in a coordinate space, is manifested by nonlocal operators. The model depends on a single dimensional constant $M$.

We find that, when the dimensional coupling constant is chosen appropriately, this model reproduces a viable background with a final stable acceleration  compatible with current constraints. Comparing our model with the one proposed in \cite{Maggiore:2014sia}, we find that the two models exhibit a different behavior in the past, but converge near the present epoch. We also observe that the background evolution of the current model is sensitive to the choice of initial conditions for the auxiliary field $W$ ($W_{0}$). In the case of vanishing $W_{0}$ the dark energy equation of state parameter $w_{\text{DE}}$ exhibits very strong phantom behavior and is in strong tension with current observational data. Furthermore, the model with vanishing $W_{0}$ does not pass the local gravity constraints on the time variation of $G$. The situation is completely different for nonvanishing choices of $W_{0}$, such that, even a small nonvanishing value of $W_{0}$ changes the overall behavior of the model sufficiently, making it compatible with observational constraints. As for any cosmological model, one should also investigate the growth of matter perturbations to ensure compatibility with observations. This is left to future work. Another work in progress is devoted to the investigation of cases with general (rational) values of the critical exponent $\nu$. These studies will allow us to deal with more realistic RG improved effective actions, where the value of critical exponent $\nu$, in general, can be arbitrary. 

\acknowledgments We thank M. Maggiore, A. Bonnano, J. Rubio and M. Pauly for reading the draft of the paper and providing us with valuable comments. We acknowledge support from DFG through
the project TRR33 ``The Dark Universe.'' H.N. acknowledges financial
support from DAAD through the program ``Forschungsstipendium f{\"u}r
Doktoranden und Nachwuchswissenschaftler'' as well as the travel support from the Heidelberg University's mobility program in the framework of the second excellence initiative.

\appendix

\section{A General Case}

\label{appendix 1}

\subsection{The model}

\noindent We consider the general model defined by the following action:
\begin{equation}
S^{\text{NL}}=\frac{1}{16\pi G}\int d^{4}x\sqrt{-g}\left[R-\mu f(R)\frac{1}{\Box^{2}}g(R)\right]+\int d^{4}x\sqrt{-g}\mathcal{L}_{\text{m}},\label{eq:fg_nlaction}
\end{equation}
where $f$ and $g$ are analytic functions of the Ricci scalar, $\mu$
stands for the scale of nonlocality and $\mathcal{L}_{\text{m}}$ is the
matter Lagrangian minimally coupled to gravity.

Taking the variation with respect to the metric tensor $g_{\mu\nu}$,
\begin{equation}
\begin{split}\delta S^{\text{NL}} & =\frac{1}{16\pi G}\int d^{4}x\delta(\sqrt{-g})\left[R-\mu f\frac{1}{\Box^{2}}g\right]\\
 & +\frac{1}{16\pi G}\int d^{4}x\sqrt{-g}\left[\delta R-\mu f,_{R}\delta R\frac{1}{\Box^{2}}g-\mu f\delta(\Box^{-2})g-\mu f\frac{1}{\Box^{2}}g,_{R}\delta R\right]+\delta\int d^{4}x\sqrt{-g}\mathcal{L}_{\text{m}}\;,
\end{split}
\label{eq:fg_nlvar}
\end{equation}
making use of the identity 
\begin{equation}
\delta(\Box^{-2})=-\Box^{-1}\delta(\Box)\Box^{-2}-\Box^{-2}\delta(\Box)\Box^{-1},\label{boxvariation}
\end{equation}
and introducing the four auxiliary fields 
\begin{align}
U & =-\Box^{-1}g\;, & Q & =-\Box^{-1}f\;,\nonumber \\
S & =-\Box^{-1}U\;, & L & =-\Box^{-1}Q\;,
\end{align}
we find the following equations of motion,
\begin{equation}
G_{\alpha\beta}=\mu K_{\alpha\beta}^{\text{NL}}+8\pi G T_{\alpha\beta},\label{eq:fgnleinstein}
\end{equation}
where the contribution coming from the nonlocal term is the following:
\begin{equation}
\begin{split}K_{\alpha\beta}^{\text{NL}}\equiv & (f,_{R}S+g,_{R}L)R_{\alpha\beta}-\nabla_{\alpha}\partial_{\beta}(f,_{R}S+g,_{R}L)+g_{\alpha\beta}\Box(f,_{R}S+g,_{R}L)+\\
 & -\frac{1}{2}g_{\alpha\beta}[fS+gL]+\frac{1}{2}g_{\alpha\beta}g^{\sigma\lambda}(\partial_{\sigma}Q\partial_{\lambda}S+\partial_{\sigma}U\partial_{\lambda}L)+\\
 & -\partial_{\alpha}U\partial_{\beta}L-\partial_{\beta}Q\partial_{\alpha}S-\frac{1}{2}g_{\alpha\beta}UQ.
\end{split}
\label{eq:fgnlktensor}
\end{equation}
For simplicity we investigate the action (\ref{eq:fg_nlaction}) in
the case when $f(R)$ and $g(R)$ have a general power-law structure.
Namely, 
and $g(R)$ are chosen to be $f(R)=R^{p}$ and $g(R)=R^{n}$, respectively,
with $n$ and $p$ being integer non-negative numbers. The equations
of motion for this model are easily obtained from Eqs.~(\ref{eq:fgnleinstein})
and (\ref{eq:fgnlktensor}) by a direct substitution 
\begin{equation}
\begin{split}G_{\alpha\beta}=\mu & \left\lbrace (R_{\alpha\beta}-\nabla_{\alpha}\nabla_{\beta}+g_{\alpha\beta}\Box)(pR^{p-1}S+nR^{n-1}L)+\right.\\
 & \left.-\frac{1}{2}g_{\alpha\beta}(R^{p}S+R^{n}L)+\frac{1}{2}g_{\alpha\beta}g^{\sigma\lambda}(\partial_{\sigma}Q\partial_{\lambda}S+\partial_{\sigma}U\partial_{\lambda}L)+\right.\\
 & \left.-\partial_{\alpha}U\partial_{\beta}L-\partial_{\beta}Q\partial_{\alpha}S-\frac{1}{2}g_{\alpha\beta}UQ\right\rbrace +8\pi GT_{\alpha\beta},
\end{split}
\label{eq:nleinstein}
\end{equation}
while the auxiliary fields satisfy the following set of coupled differential
equations 
\begin{align}\label{auxgen}
\Box U & =-R^{n}\;, & \Box Q & =-R^{p}\;,\nonumber \\
\Box S & =-U\;, & \Box L & =-Q\;.
\end{align}

If we introduce in the Lagrangian a parameter $M$ with the dimension
of a mass, we can redefine the $\mu$ parameter in \eqref{eq:nleinstein}
as follows: 
\begin{equation}
\mu=\frac{1}{6}M^{-2(n+p)+6}\;.\label{muparam}
\end{equation}
If $n=1$ and $p=1$ from Eqs.~(\ref{eq:nleinstein}) and (\ref{auxgen})
we find the equations of motion of $RR$ nonlocal gravity model~\cite{Maggiore:2014sia}.
\begin{align}
G_{\alpha\beta} & =\frac{M^{2}}{6}K_{\alpha\beta}^{\text{RR}}+8\pi GT_{\alpha\beta},\label{eq:RReinstein}\\
\Box U & =-R,\label{eq:RRudef}\\
\Box S & =-U,\label{eq:RRsdef}
\end{align}
where $K_{\alpha\beta}^{\text{RR}}$ tensor, which stands for the correction to Einstein equations coming from nonlocal corrections, is defined as
\begin{equation}
\begin{split}K_{\alpha\beta}^{\text{RR}}\equiv & \quad2(G_{\alpha\beta}-\nabla_{\alpha}\nabla_{\beta}+g_{\alpha\beta}\Box)S+g_{\alpha\beta}g^{\sigma\lambda}\nabla_{\sigma}U\nabla_{\lambda}S+\\
 & -(\nabla_{\alpha}U\nabla_{\beta}S+\nabla_{\beta}U\nabla_{\alpha}S)-\frac{1}{2}g_{\alpha\beta}U^{2}.
\end{split}
\end{equation}
These equations fully coincide with those for the $RR$ nonlocal gravity model~\cite{Maggiore:2014sia}.

\subsection{Cosmological equations}

In this section we will study the background cosmology of our model.
To do this we choose our metric to be of a flat FLRW type 
\begin{equation}
ds^{2}=-\frac{1}{H^{2}}dN^{2}+a^{2}d\vec{x}^{2},\label{eq:flrwmetric}
\end{equation}
with $a$ being the scale factor, $H$ the Hubble rate and $N=\log a$
the number of e-foldings.

We also introduce the following quantities 
\begin{equation}
h=\frac{H}{H_{0}},\hspace{10mm}\xi\equiv\frac{H'}{H}=\frac{h'}{h},\label{eq:xidef}
\end{equation}
where $H_{0}$ is the Hubble rate today.
For the cosmological analysis it is sometimes useful to go from dimensionful quantities to dimensionless ones. To do this we multiply our dimensionful auxiliary fields by powers of $H_{0}$ and as such we define new dimensionless auxiliary fields as follows

\begin{align*}
X & \equiv H_{0}^{2-2n}U, & W & \equiv H_{0}^{2-2p}Q,\\
V & \equiv H_{0}^{4-2n}S, & Z & \equiv H_{0}^{4-2p}L.
\end{align*}
\noindent From the $(00)$ component of Eq. (\ref{eq:nleinstein})
one can express $h^{2}$ through new dimensionless functions defined above:
\begin{equation}
h^{2}=\frac{2\mu}{3H_{0}^{6-2(n+p)}}h^{2}Y^{\text{NL}}+\frac{8\pi G}{3H_{0}^{2}}\rho,\label{eq:nleinstein2_zero}
\end{equation}
where all terms coming from the nonlocal part are collected in the
following quantity: 
\begin{equation}
\begin{split}Y^{\text{NL}}= & \frac{1}{4h^{2}}\left\lbrace WX-h^{2}\left(V'W'+Z'X'\right)+h^{2p}\left[C_{p}\left(VB_{p}+pV'\right)+D_{p}V\xi'\right]+\right.\\
 & \left.+h^{2n}\left[C_{n}(ZB_{n}+nZ')+D_{n}Z\xi'\right]\right\rbrace,
\end{split}
\label{eq:nly_mn}
\end{equation}
where, to simplify the notation, we have defined the following coefficients
\begin{align}
B_{k} & \equiv(2k-1)(k-1)\xi-k+2\;,\label{eq:bdef}\\
C_{k} & \equiv6^{k}(\xi+2)^{k-1}\;,\label{eq:cdef}\\
D_{k} & \equiv k(k-1)6^{k}(\xi+2)^{k-2}\;.\label{eq:ddef}
\end{align}
These terms only depends on $\xi$ and on $k$, where the latter takes
the values of either $n$ or $p$. We will also express the dimensionful scale parameter of the model $\mu$ through a new dimensionless quantity defined as
\begin{equation}
\gamma\equiv \frac{2}{3} \mu H_{0}^{2(n+p)-6}\;.\label{defgamma}
\end{equation}
\bibliographystyle{apsrev}
\bibliography{amendolamodnonloc,Henrik,nonlocalRefsTSK}

\begin{thebibliography}{35}
\expandafter\ifx\csname natexlab\endcsname\relax\def\natexlab#1{#1}\fi
\expandafter\ifx\csname bibnamefont\endcsname\relax
  \def\bibnamefont#1{#1}\fi
\expandafter\ifx\csname bibfnamefont\endcsname\relax
  \def\bibfnamefont#1{#1}\fi
\expandafter\ifx\csname citenamefont\endcsname\relax
  \def\citenamefont#1{#1}\fi
\expandafter\ifx\csname url\endcsname\relax
  \def\url#1{\texttt{#1}}\fi
\expandafter\ifx\csname urlprefix\endcsname\relax\def\urlprefix{URL }\fi
\providecommand{\bibinfo}[2]{#2}
\providecommand{\eprint}[2][]{\url{#2}}

\bibitem[{\citenamefont{Riess et~al.}(1998)}]{Riess:1998cb}
\bibinfo{author}{\bibfnamefont{A.~G.} \bibnamefont{Riess}} \bibnamefont{et~al.}
  (\bibinfo{collaboration}{Supernova Search Team}), \bibinfo{journal}{Astron.
  J.} \textbf{\bibinfo{volume}{116}}, \bibinfo{pages}{1009}
  (\bibinfo{year}{1998}), \eprint{astro-ph/9805201}.

\bibitem[{\citenamefont{Perlmutter et~al.}(1999)}]{Perlmutter:1998np}
\bibinfo{author}{\bibfnamefont{S.}~\bibnamefont{Perlmutter}}
  \bibnamefont{et~al.} (\bibinfo{collaboration}{Supernova Cosmology Project}),
  \bibinfo{journal}{Astrophys. J.} \textbf{\bibinfo{volume}{517}},
  \bibinfo{pages}{565} (\bibinfo{year}{1999}), \eprint{astro-ph/9812133}.

\bibitem[{\citenamefont{Sherwin et~al.}(2011)}]{Sherwin:2011gv}
\bibinfo{author}{\bibfnamefont{B.~D.} \bibnamefont{Sherwin}}
  \bibnamefont{et~al.}, \bibinfo{journal}{Phys. Rev. Lett.}
  \textbf{\bibinfo{volume}{107}}, \bibinfo{pages}{021302}
  (\bibinfo{year}{2011}), \eprint{1105.0419}.

\bibitem[{\citenamefont{Dunkley et~al.}(2009)}]{Dunkley:2008ie}
\bibinfo{author}{\bibfnamefont{J.}~\bibnamefont{Dunkley}} \bibnamefont{et~al.}
  (\bibinfo{collaboration}{WMAP}), \bibinfo{journal}{Astrophys. J. Suppl.}
  \textbf{\bibinfo{volume}{180}}, \bibinfo{pages}{306} (\bibinfo{year}{2009}),
  \eprint{0803.0586}.

\bibitem[{\citenamefont{Komatsu et~al.}(2009)}]{Komatsu:2008hk}
\bibinfo{author}{\bibfnamefont{E.}~\bibnamefont{Komatsu}} \bibnamefont{et~al.}
  (\bibinfo{collaboration}{WMAP}), \bibinfo{journal}{Astrophys. J. Suppl.}
  \textbf{\bibinfo{volume}{180}}, \bibinfo{pages}{330} (\bibinfo{year}{2009}),
  \eprint{0803.0547}.

\bibitem[{\citenamefont{van Engelen et~al.}(2012)}]{vanEngelen:2012va}
\bibinfo{author}{\bibfnamefont{A.}~\bibnamefont{van Engelen}}
  \bibnamefont{et~al.}, \bibinfo{journal}{Astrophys. J.}
  \textbf{\bibinfo{volume}{756}}, \bibinfo{pages}{142} (\bibinfo{year}{2012}),
  \eprint{1202.0546}.

\bibitem[{\citenamefont{Scranton et~al.}(2003)}]{Scranton:2003in}
\bibinfo{author}{\bibfnamefont{R.}~\bibnamefont{Scranton}} \bibnamefont{et~al.}
  (\bibinfo{collaboration}{SDSS}) (\bibinfo{year}{2003}),
  \eprint{astro-ph/0307335}.

\bibitem[{\citenamefont{Sanchez et~al.}(2012)}]{Sanchez:2012sg}
\bibinfo{author}{\bibfnamefont{A.~G.} \bibnamefont{Sanchez}}
  \bibnamefont{et~al.}, \bibinfo{journal}{Mon. Not. Roy. Astron. Soc.}
  \textbf{\bibinfo{volume}{425}}, \bibinfo{pages}{415} (\bibinfo{year}{2012}),
  \eprint{1203.6616}.

\bibitem[{\citenamefont{Wetterich}(2017)}]{Wetterich:2017ixo}
\bibinfo{author}{\bibfnamefont{C.}~\bibnamefont{Wetterich}}
  (\bibinfo{year}{2017}), \eprint{1704.08040}.

\bibitem[{\citenamefont{Hamber and Williams}(2005)}]{Hamber:2005dw}
\bibinfo{author}{\bibfnamefont{H.~W.} \bibnamefont{Hamber}} \bibnamefont{and}
  \bibinfo{author}{\bibfnamefont{R.~M.} \bibnamefont{Williams}},
  \bibinfo{journal}{Phys. Rev.} \textbf{\bibinfo{volume}{D72}},
  \bibinfo{pages}{044026} (\bibinfo{year}{2005}), \eprint{hep-th/0507017}.

\bibitem[{\citenamefont{Hamber and Williams}(2007)}]{Hamber:2006sv}
\bibinfo{author}{\bibfnamefont{H.~W.} \bibnamefont{Hamber}} \bibnamefont{and}
  \bibinfo{author}{\bibfnamefont{R.~M.} \bibnamefont{Williams}},
  \bibinfo{journal}{Phys. Rev.} \textbf{\bibinfo{volume}{D75}},
  \bibinfo{pages}{084014} (\bibinfo{year}{2007}), \eprint{hep-th/0607228}.

\bibitem[{\citenamefont{Bonanno and Saueressig}(2017)}]{Bonanno:2017pkg}
\bibinfo{author}{\bibfnamefont{A.}~\bibnamefont{Bonanno}} \bibnamefont{and}
  \bibinfo{author}{\bibfnamefont{F.}~\bibnamefont{Saueressig}}
  (\bibinfo{year}{2017}), \eprint{1702.04137}.

\bibitem[{\citenamefont{Bonanno and Platania}(2015)}]{Bonanno:2015fga}
\bibinfo{author}{\bibfnamefont{A.}~\bibnamefont{Bonanno}} \bibnamefont{and}
  \bibinfo{author}{\bibfnamefont{A.}~\bibnamefont{Platania}},
  \bibinfo{journal}{Phys. Lett.} \textbf{\bibinfo{volume}{B750}},
  \bibinfo{pages}{638} (\bibinfo{year}{2015}), \eprint{1507.03375}.

\bibitem[{\citenamefont{Maggiore and Mancarella}(2014)}]{Maggiore:2014sia}
\bibinfo{author}{\bibfnamefont{M.}~\bibnamefont{Maggiore}} \bibnamefont{and}
  \bibinfo{author}{\bibfnamefont{M.}~\bibnamefont{Mancarella}},
  \bibinfo{journal}{Phys. Rev.} \textbf{\bibinfo{volume}{D90}},
  \bibinfo{pages}{023005} (\bibinfo{year}{2014}), \eprint{1402.0448}.

\bibitem[{\citenamefont{Hamber and Williams}(1995)}]{Hamber:1994jh}
\bibinfo{author}{\bibfnamefont{H.~W.} \bibnamefont{Hamber}} \bibnamefont{and}
  \bibinfo{author}{\bibfnamefont{R.~M.} \bibnamefont{Williams}},
  \bibinfo{journal}{Nucl. Phys.} \textbf{\bibinfo{volume}{B435}},
  \bibinfo{pages}{361} (\bibinfo{year}{1995}), \eprint{hep-th/9406163}.

\bibitem[{\citenamefont{Hamber and Williams}(2004)}]{Hamber:2004ew}
\bibinfo{author}{\bibfnamefont{H.~W.} \bibnamefont{Hamber}} \bibnamefont{and}
  \bibinfo{author}{\bibfnamefont{R.~M.} \bibnamefont{Williams}},
  \bibinfo{journal}{Phys. Rev.} \textbf{\bibinfo{volume}{D70}},
  \bibinfo{pages}{124007} (\bibinfo{year}{2004}), \eprint{hep-th/0407039}.

\bibitem[{\citenamefont{Litim}(2004)}]{Litim:2003vp}
\bibinfo{author}{\bibfnamefont{D.~F.} \bibnamefont{Litim}},
  \bibinfo{journal}{Phys. Rev. Lett.} \textbf{\bibinfo{volume}{92}},
  \bibinfo{pages}{201301} (\bibinfo{year}{2004}), \eprint{hep-th/0312114}.

\bibitem[{\citenamefont{Brezin et~al.}(1976)\citenamefont{Brezin, Zinn-Justin,
  and Le~Guillou}}]{Brezin:1976ap}
\bibinfo{author}{\bibfnamefont{E.}~\bibnamefont{Brezin}},
  \bibinfo{author}{\bibfnamefont{J.}~\bibnamefont{Zinn-Justin}},
  \bibnamefont{and} \bibinfo{author}{\bibfnamefont{J.~C.}
  \bibnamefont{Le~Guillou}}, \bibinfo{journal}{Phys. Rev.}
  \textbf{\bibinfo{volume}{D14}}, \bibinfo{pages}{2615} (\bibinfo{year}{1976}).

\bibitem[{\citenamefont{Reuter and Saueressig}(2002)}]{Reuter:2001ag}
\bibinfo{author}{\bibfnamefont{M.}~\bibnamefont{Reuter}} \bibnamefont{and}
  \bibinfo{author}{\bibfnamefont{F.}~\bibnamefont{Saueressig}},
  \bibinfo{journal}{Phys. Rev.} \textbf{\bibinfo{volume}{D65}},
  \bibinfo{pages}{065016} (\bibinfo{year}{2002}), \eprint{hep-th/0110054}.

\bibitem[{\citenamefont{Barnaby and Kamran}(2008)}]{Barnaby:2007ve}
\bibinfo{author}{\bibfnamefont{N.}~\bibnamefont{Barnaby}} \bibnamefont{and}
  \bibinfo{author}{\bibfnamefont{N.}~\bibnamefont{Kamran}},
  \bibinfo{journal}{JHEP} \textbf{\bibinfo{volume}{02}}, \bibinfo{pages}{008}
  (\bibinfo{year}{2008}), \eprint{0709.3968}.

\bibitem[{\citenamefont{Lopez~Nacir and Mazzitelli}(2007)}]{LopezNacir:2006tn}
\bibinfo{author}{\bibfnamefont{D.}~\bibnamefont{Lopez~Nacir}} \bibnamefont{and}
  \bibinfo{author}{\bibfnamefont{F.~D.} \bibnamefont{Mazzitelli}},
  \bibinfo{journal}{Phys. Rev.} \textbf{\bibinfo{volume}{D75}},
  \bibinfo{pages}{024003} (\bibinfo{year}{2007}), \eprint{hep-th/0610031}.

\bibitem[{\citenamefont{Vardanyan et~al.}(2017)\citenamefont{Vardanyan, Akrami,
  Amendola, and Silvestri}}]{Vardanyan:2017kal}
\bibinfo{author}{\bibfnamefont{V.}~\bibnamefont{Vardanyan}},
  \bibinfo{author}{\bibfnamefont{Y.}~\bibnamefont{Akrami}},
  \bibinfo{author}{\bibfnamefont{L.}~\bibnamefont{Amendola}}, \bibnamefont{and}
  \bibinfo{author}{\bibfnamefont{A.}~\bibnamefont{Silvestri}}
  (\bibinfo{year}{2017}), \eprint{1702.08908}.

\bibitem[{\citenamefont{Nersisyan et~al.}(2016)\citenamefont{Nersisyan, Akrami,
  Amendola, Koivisto, and Rubio}}]{Nersisyan:2016hjh}
\bibinfo{author}{\bibfnamefont{H.}~\bibnamefont{Nersisyan}},
  \bibinfo{author}{\bibfnamefont{Y.}~\bibnamefont{Akrami}},
  \bibinfo{author}{\bibfnamefont{L.}~\bibnamefont{Amendola}},
  \bibinfo{author}{\bibfnamefont{T.~S.} \bibnamefont{Koivisto}},
  \bibnamefont{and} \bibinfo{author}{\bibfnamefont{J.}~\bibnamefont{Rubio}},
  \bibinfo{journal}{Phys. Rev.} \textbf{\bibinfo{volume}{D94}},
  \bibinfo{pages}{043531} (\bibinfo{year}{2016}), \eprint{1606.04349}.

\bibitem[{\citenamefont{Ade et~al.}(2016)}]{Ade:2015rim}
\bibinfo{author}{\bibfnamefont{P.~A.~R.} \bibnamefont{Ade}}
  \bibnamefont{et~al.} (\bibinfo{collaboration}{Planck}),
  \bibinfo{journal}{Astron. Astrophys.} \textbf{\bibinfo{volume}{594}},
  \bibinfo{pages}{A14} (\bibinfo{year}{2016}), \eprint{1502.01590}.

\bibitem[{\citenamefont{Suzuki et~al.}(2012)}]{Suzuki:2011hu}
\bibinfo{author}{\bibfnamefont{N.}~\bibnamefont{Suzuki}} \bibnamefont{et~al.},
  \bibinfo{journal}{Astrophys. J.} \textbf{\bibinfo{volume}{746}},
  \bibinfo{pages}{85} (\bibinfo{year}{2012}), \eprint{1105.3470}.

\bibitem[{\citenamefont{Maggiore}(2016{\natexlab{a}})}]{Maggiore:2016gpx}
\bibinfo{author}{\bibfnamefont{M.}~\bibnamefont{Maggiore}}
  (\bibinfo{year}{2016}{\natexlab{a}}), \eprint{1606.08784}.

\bibitem[{\citenamefont{Maggiore}(2016{\natexlab{b}})}]{Maggiore:2016fbn}
\bibinfo{author}{\bibfnamefont{M.}~\bibnamefont{Maggiore}},
  \bibinfo{journal}{Phys. Rev.} \textbf{\bibinfo{volume}{D93}},
  \bibinfo{pages}{063008} (\bibinfo{year}{2016}{\natexlab{b}}),
  \eprint{1603.01515}.

\bibitem[{\citenamefont{Williams et~al.}(2004)\citenamefont{Williams, Turyshev,
  and Boggs}}]{Williams:2004qba}
\bibinfo{author}{\bibfnamefont{J.~G.} \bibnamefont{Williams}},
  \bibinfo{author}{\bibfnamefont{S.~G.} \bibnamefont{Turyshev}},
  \bibnamefont{and} \bibinfo{author}{\bibfnamefont{D.~H.} \bibnamefont{Boggs}},
  \bibinfo{journal}{Phys. Rev. Lett.} \textbf{\bibinfo{volume}{93}},
  \bibinfo{pages}{261101} (\bibinfo{year}{2004}), \eprint{gr-qc/0411113}.

\bibitem[{\citenamefont{Barreira et~al.}(2014)\citenamefont{Barreira, Li,
  Hellwing, Baugh, and Pascoli}}]{Barreira:2014kra}
\bibinfo{author}{\bibfnamefont{A.}~\bibnamefont{Barreira}},
  \bibinfo{author}{\bibfnamefont{B.}~\bibnamefont{Li}},
  \bibinfo{author}{\bibfnamefont{W.~A.} \bibnamefont{Hellwing}},
  \bibinfo{author}{\bibfnamefont{C.~M.} \bibnamefont{Baugh}}, \bibnamefont{and}
  \bibinfo{author}{\bibfnamefont{S.}~\bibnamefont{Pascoli}},
  \bibinfo{journal}{JCAP} \textbf{\bibinfo{volume}{1409}}, \bibinfo{pages}{031}
  (\bibinfo{year}{2014}), \eprint{1408.1084}.

\bibitem[{\citenamefont{Dirian et~al.}(2016)\citenamefont{Dirian, Foffa, Kunz,
  Maggiore, and Pettorino}}]{Dirian:2016puz}
\bibinfo{author}{\bibfnamefont{Y.}~\bibnamefont{Dirian}},
  \bibinfo{author}{\bibfnamefont{S.}~\bibnamefont{Foffa}},
  \bibinfo{author}{\bibfnamefont{M.}~\bibnamefont{Kunz}},
  \bibinfo{author}{\bibfnamefont{M.}~\bibnamefont{Maggiore}}, \bibnamefont{and}
  \bibinfo{author}{\bibfnamefont{V.}~\bibnamefont{Pettorino}},
  \bibinfo{journal}{JCAP} \textbf{\bibinfo{volume}{1605}}, \bibinfo{pages}{068}
  (\bibinfo{year}{2016}), \eprint{1602.03558}.

\bibitem[{\citenamefont{Dirian}(2017)}]{Dirian:2017pwp}
\bibinfo{author}{\bibfnamefont{Y.}~\bibnamefont{Dirian}}
  (\bibinfo{year}{2017}), \eprint{1704.04075}.

\bibitem[{\citenamefont{Nersisyan
  et~al.}(2017{\natexlab{a}})\citenamefont{Nersisyan, Cid, and
  Amendola}}]{Nersisyan:2017mgj}
\bibinfo{author}{\bibfnamefont{H.}~\bibnamefont{Nersisyan}},
  \bibinfo{author}{\bibfnamefont{A.~F.} \bibnamefont{Cid}}, \bibnamefont{and}
  \bibinfo{author}{\bibfnamefont{L.}~\bibnamefont{Amendola}},
  \bibinfo{journal}{JCAP} \textbf{\bibinfo{volume}{1704}}, \bibinfo{pages}{046}
  (\bibinfo{year}{2017}{\natexlab{a}}), \eprint{1701.00434}.

\bibitem[{\citenamefont{Nersisyan
  et~al.}(2017{\natexlab{b}})\citenamefont{Nersisyan, Akrami, Amendola,
  Koivisto, Rubio, and Solomon}}]{Nersisyan:2016jta}
\bibinfo{author}{\bibfnamefont{H.}~\bibnamefont{Nersisyan}},
  \bibinfo{author}{\bibfnamefont{Y.}~\bibnamefont{Akrami}},
  \bibinfo{author}{\bibfnamefont{L.}~\bibnamefont{Amendola}},
  \bibinfo{author}{\bibfnamefont{T.~S.} \bibnamefont{Koivisto}},
  \bibinfo{author}{\bibfnamefont{J.}~\bibnamefont{Rubio}}, \bibnamefont{and}
  \bibinfo{author}{\bibfnamefont{A.~R.} \bibnamefont{Solomon}},
  \bibinfo{journal}{Phys. Rev.} \textbf{\bibinfo{volume}{D95}},
  \bibinfo{pages}{043539} (\bibinfo{year}{2017}{\natexlab{b}}),
  \eprint{1610.01799}.

\bibitem[{\citenamefont{Foffa et~al.}(2014)\citenamefont{Foffa, Maggiore, and
  Mitsou}}]{Foffa:2013vma}
\bibinfo{author}{\bibfnamefont{S.}~\bibnamefont{Foffa}},
  \bibinfo{author}{\bibfnamefont{M.}~\bibnamefont{Maggiore}}, \bibnamefont{and}
  \bibinfo{author}{\bibfnamefont{E.}~\bibnamefont{Mitsou}},
  \bibinfo{journal}{Int. J. Mod. Phys.} \textbf{\bibinfo{volume}{A29}},
  \bibinfo{pages}{1450116} (\bibinfo{year}{2014}), \eprint{1311.3435}.

\bibitem[{\citenamefont{Dirian et~al.}(2014)\citenamefont{Dirian, Foffa,
  Khosravi, Kunz, and Maggiore}}]{Dirian:2014ara}
\bibinfo{author}{\bibfnamefont{Y.}~\bibnamefont{Dirian}},
  \bibinfo{author}{\bibfnamefont{S.}~\bibnamefont{Foffa}},
  \bibinfo{author}{\bibfnamefont{N.}~\bibnamefont{Khosravi}},
  \bibinfo{author}{\bibfnamefont{M.}~\bibnamefont{Kunz}}, \bibnamefont{and}
  \bibinfo{author}{\bibfnamefont{M.}~\bibnamefont{Maggiore}},
  \bibinfo{journal}{JCAP} \textbf{\bibinfo{volume}{1406}}, \bibinfo{pages}{033}
  (\bibinfo{year}{2014}), \eprint{1403.6068}.

\end{thebibliography}
 
\end{document}